\documentclass{article}

    \PassOptionsToPackage{numbers, compress}{natbib}
  \usepackage[preprint]{neurips_2026}

\usepackage[utf8]{inputenc} % allow utf-8 input
\usepackage[T1]{fontenc}    % use 8-bit T1 fonts
\usepackage{hyperref}       % hyperlinks
\usepackage{url}            % simple URL typesetting
\usepackage{booktabs}       % professional-quality tables
\usepackage{amsfonts}       % blackboard math symbols
\usepackage{nicefrac}       % compact symbols for 1/2, etc.
\usepackage{microtype}      % microtypography
\usepackage{xcolor}         % colors

\usepackage[mathscr]{eucal}
\usepackage{amssymb,amsmath,amsfonts,latexsym}
\usepackage{xcolor,url,overpic}
\usepackage{array}%array and tabular environments
\usepackage{verbatim}
\usepackage{bm}
\usepackage{bbm}
\usepackage{dsfont}
\usepackage{algorithmic}
\usepackage{algorithm}
\usepackage{verbatim}
\usepackage{textcomp}
\usepackage{mathrsfs}
\usepackage{epstopdf}
\usepackage{booktabs} % for professional tables
\usepackage{thm-restate}

\newcommand{\openone}{\leavevmode\hbox{\small1\normalsize\kern-.33em1}}

\catcode`~=11 \def\UrlSpecials{\do\~{\kern -.15em\lower .7ex\hbox{~}\kern .04em}} \catcode`~=13 

\allowdisplaybreaks[1]

\newcommand{\nn}{\nonumber}

\newcommand{\calA}{\mathcal{A}}
\newcommand{\calB}{\mathcal{B}}

\newcommand{\calG}{\mathcal{G}}

\newcommand{\calM}{\mathcal{M}}

\newcommand{\calP}{\mathcal{P}}

\newcommand{\calR}{\mathcal{R}}
\newcommand{\calS}{\mathcal{S}}

\newcommand{\calV}{\mathcal{V}}

\newcommand{\calX}{\mathcal{X}}

\newcommand{\calZ}{\mathcal{Z}}

\newcommand{\bv}{\mathbf{v}}

\newcommand{\rmc}{\mathrm{c}}

\newcommand{\rmH}{\mathrm{H}}

\newcommand{\bbE}{\mathbb{E}}

\newcommand{\bbN}{\mathbb{N}}

\newcommand{\bbP}{\mathbb{P}}

\newcommand{\bbR}{\mathbb{R}}

\newcommand{\bbZ}{\mathbb{Z}}

\DeclareMathAlphabet{\mathbsf}{OT1}{cmss}{bx}{n}
\DeclareMathAlphabet{\mathssf}{OT1}{cmss}{m}{sl}% slanted sans serif

\newcommand{\rvD}{\mathsf{D}}

\newcommand{\rvH}{\mathsf{H}}

\newcommand{\rvI}{\mathsf{I}}

\DeclareSymbolFont{bsfletters}{OT1}{cmss}{bx}{n}  
\DeclareSymbolFont{ssfletters}{OT1}{cmss}{m}{n}
\DeclareMathSymbol{\bsfGamma}{0}{bsfletters}{'000}
\DeclareMathSymbol{\ssfGamma}{0}{ssfletters}{'000}
\DeclareMathSymbol{\bsfDelta}{0}{bsfletters}{'001}
\DeclareMathSymbol{\ssfDelta}{0}{ssfletters}{'001}
\DeclareMathSymbol{\bsfTheta}{0}{bsfletters}{'002}
\DeclareMathSymbol{\ssfTheta}{0}{ssfletters}{'002}
\DeclareMathSymbol{\bsfLambda}{0}{bsfletters}{'003}
\DeclareMathSymbol{\ssfLambda}{0}{ssfletters}{'003}
\DeclareMathSymbol{\bsfXi}{0}{bsfletters}{'004}
\DeclareMathSymbol{\ssfXi}{0}{ssfletters}{'004}
\DeclareMathSymbol{\bsfPi}{0}{bsfletters}{'005}
\DeclareMathSymbol{\ssfPi}{0}{ssfletters}{'005}
\DeclareMathSymbol{\bsfSigma}{0}{bsfletters}{'006}
\DeclareMathSymbol{\ssfSigma}{0}{ssfletters}{'006}
\DeclareMathSymbol{\bsfUpsilon}{0}{bsfletters}{'007}
\DeclareMathSymbol{\ssfUpsilon}{0}{ssfletters}{'007}
\DeclareMathSymbol{\bsfPhi}{0}{bsfletters}{'010}
\DeclareMathSymbol{\ssfPhi}{0}{ssfletters}{'010}
\DeclareMathSymbol{\bsfPsi}{0}{bsfletters}{'011}
\DeclareMathSymbol{\ssfPsi}{0}{ssfletters}{'011}
\DeclareMathSymbol{\bsfOmega}{0}{bsfletters}{'012}
\DeclareMathSymbol{\ssfOmega}{0}{ssfletters}{'012}

\newcommand{\hatM}{\hat{M}}

\newcommand{\dotleq}{\stackrel{.}{\leq}}

\newcommand{\eqb}{\stackrel{(b)}{=}}

\DeclareMathOperator*{\argmax}{arg\,max}
\DeclareMathOperator*{\argmin}{arg\,min}

\usepackage{thmtools, thm-restate}
\newtheorem{theorem}{Theorem} 
\newtheorem{lemma}[theorem]{Lemma}

\newtheorem{definition}{Definition}

\newtheorem{remark}{Remark}

\newtheorem{assumption}{Assumption}

\usepackage{times}
\usepackage{mathtools}
\usepackage{cite}
\usepackage{graphicx}
\usepackage{enumitem}
\mathtoolsset{showonlyrefs}
\usepackage[caption=false]{subfig}

\title{Fundamental Trade-Offs in Multi-Bit Watermarking of Stochastic Processes}

\author{%
  Haiyun He \\
  HKUST (GZ)\\
  \texttt{haiyunhe@hkust-gz.edu.cn} \\
  \And
  Yepeng Liu \\
  UC Santa Barbara \\
  \texttt{yepengliu@ucsb.edu} \\
  \And
  Zhuoer Shen \\
  UC Santa Barbara \\
  \texttt{zhuoershen@ucsb.edu} \\
  \And
  Ziqiao Wang \\
  Tongji University \\
  \texttt{ziqiaowang@tongji.edu.cn} \\
  \And
  Yongyi Mao \\
  University of Ottawa \\
  \texttt{ymao@uottawa.ca} \\
  \And
  Yuheng Bu \\
  UC Santa Barbara \\
  \texttt{buyuheng@ucsb.edu} \\
}

\newcommand{\KL}{\mathsf{KL}}
\newcommand{\TV}{\mathsf{TV}}
\newcommand{\LB}{\mathsf{LB}}

\newcommand{\hh}[1]{{\color{red}[HH: #1]}}

\newcommand{\key}{\mathtt{key}}

\begin{document}

\maketitle

\begin{abstract}%
We study multi-bit watermarking for data generated by stochastic processes, where a hidden message is embedded during sampling and must be decodable by an authorized detector that possesses side information unavailable to unauthorized observers. In high-stakes deployments, a practical watermark must simultaneously control false alarms, preserve generation quality without distorting the output distribution, and support reliable multi-bit decoding. Satisfying all three goals at once inevitably creates fundamental trade-offs. We formulate watermark embedding as a distributional information-embedding problem and watermark detection as a multiple-hypothesis testing problem under distortion and rate constraints, leading to four fundamental metrics: false-alarm probability, detection error probability, distortion, and information rate. Within this information-theoretic framework, we derive matched converse and achievability bounds that characterize the optimal trade-offs and provide scheme-agnostic benchmarks for any watermarking method. For stationary ergodic stochastic processes, we further obtain matched asymptotic limits and connect them to the finite-sample regime. Finally, we present a reference watermarking construction satisfying our assumptions and empirically illustrating the predicted trade-offs.
\end{abstract}

\section{Introduction}

Many modern systems generate data by sampling from stochastic processes specified by a known or learned generative distribution. This umbrella includes classical Markov~\citep{davis2018markov} and autoregressive models, contemporary diffusion models~\citep{song2020score}, and large language models (LLMs)~\citep{achiam2023gpt}, and extends to protein/DNA sequence generators~\citep{madani2020progen}, randomized simulators~\citep{tran2017hierarchical}, and synthetic data generation~\citep{lu2023machine}.

Within such stochastic systems, it is increasingly desirable to embed hidden information directly into the generation process so that each generated sample carries verifiable watermarks. The embedded watermarks can certify ownership by identifying the source model or provider, or establish provenance by recording the generation context or other traceable identifiers. It can also support integrity guarantees by making tampering or unauthorized post-processing easier to detect. This goal naturally generalizes existing zero-bit watermarking in generative models~\citep{kirchenbauer2023watermark,zhao2023provable,liu2024adaptive, dathathri2024scalable,kuditipudi2023robust}, which only enables detection of AI-generated content, to multi-bit watermarking~\citep{yoo2024advancing,cohen2025watermarking,qu2025provably,huang2026optimalmultibit}, where an authorized detector can not only detect the presence of a watermark but also decode an embedded message. For a more comprehensive literature review, please refer to Appendix \ref{Sec: related works}.
Despite notable recent advances in watermarking for generative AI~\citep{yang2025watermarking}, much of the existing literature remains ad hoc and empirical, and 
%lacks a principled understanding of the fundamental trade-offs that govern multi-bit watermarking design.
lacks in-depth understanding of the tension between the various performance metrics and how they trade off at a fundamental level. 

These trade-offs are especially critical in high-stakes settings, where watermarking must satisfy multiple, often competing, design requirements. Watermarking must tightly control false-alarms to avoid costly misattribution, imposing stringent false positive rate (FPR) constraints on detection~\citep{he2025distributional}. At the same time, it should be distortion-free: watermarked outputs must preserve generation quality and remain statistically indistinguishable from unwatermarked ones~\citep{boroujeny2024multi}. For multi-bit watermarking, this distribution-preservation requirement must hold uniformly over the message space, meaning that for any embedded message, the output distribution should match the unwatermarked distribution. However, reliable multi-bit decoding pushes in the opposite direction, because increasing payload inevitably perturbs the generative distribution. Therefore, reconciling low false-alarms, reliable multi-bit decoding, and distribution-preserving generation is what makes these trade-offs fundamental. This motivates a foundational question: \emph{how much information can be embedded and reliably decoded from a stochastic process while keeping the resulting distribution indistinguishable from the original?}

This paper formulates multi-bit watermarking of stochastic processes as a distributional information embedding problem. We study how to embed a message into samples from a generative distribution so that an authorized detector can reliably decode it with side information (e.g., secret key), while the resulting outputs remain statistically indistinguishable from unwatermarked samples. The framework abstracts away model-specific details and applies broadly to autoregressive and Markov processes, including language generation and protein sequence generation. 

% By casting watermarking as inference on stochastic processes, we characterize fundamental limits on detectability/decodability, distortion-free (distribution-preserving) generation, and embedding rate.

This viewpoint yields a clean formulation of watermark design as constrained optimization: maximize detection and decoding performance subject to worst-case false-alarm constraint and generation distortion. We summarize performance through four facets $(\alpha,\frac{1}{m}\sum_{j=1}^m \beta_j, d, R)$: the false-alarm probability $\alpha$, the average detection error $\frac{1}{m}\sum_{j=1}^m \beta_j$, the distortion level $d$ (deviation from the unwatermarked process), and the information rate $R$ (average payload per sample). Framing watermarking in information-theoretic terms puts these objectives in a common language and makes the unavoidable trade-offs explicit. Our results establish an information-theoretic benchmark that any multi-bit watermarking scheme must satisfy, regardless of its
embedding mechanism.

Our contribution includes:
\begin{itemize}
[itemsep=0pt,topsep=0pt,leftmargin=12pt]
    \item We formulate multi-bit watermark embedding as a distributional information embedding problem and watermark detection as a hypothesis test performed by an authorized detector with side information, under distortion and rate constraints. This yields four fundamental performance metrics: false-alarm probability, detection error probability, distortion, and embedded information rate.
    \item We derive matched information-theoretic converse and achievability bounds that characterize the fundamental trade-offs among these metrics, establishing a performance benchmark applicable to any watermarking scheme of a stochastic process within our framework.
    \item For stationary ergodic stochastic processes, we develop an asymptotic characterization with matched upper and lower bounds, and show how the finite-sample trade-offs converge to the asymptotic limits.
    
    \item We propose a reference watermarking construction that satisfies the theoretical assumptions and empirically demonstrates the predicted trade-offs, serving as a concrete instantiation of the theory. Numerical results are presented to visualize the trade-offs and the achievable region.  
    % \hh{Markov chain: bound and achievability}
    % \item Our results provide a principled evaluation framework that clarifies how watermarking performance should be compared and reported in future work.
\end{itemize}

\section{Problem Formulation}
\subsection{Notation}
For any set $\calX$, we denote the space of all probability measures over  $\calX$ by $\calP(\calX)$. For a random variable taking values in $\calX$, we use $P_X$ or $Q_X$ to denote its distribution and use the lower-cased letter $x$ to denote a realization of $X$. 
For a sequence of random variables $X_1, \ldots, X_n$, and any $i, j \in [n]$ with  $i\le j$, we denote $X_i^j\coloneqq (X_i,\ldots,X_j)$ and drop the subscript if $i=1$. 
%Denote $X\rvIm P_X\in\cP(\cX)$ as a random variable, and let $x\in\cX$ be the realization of $X$. 
We may use distortion function, namely, a function $\rvD: \calP(\calX) \times \calP(\calX) \to [0,+\infty)$
to measure the dissimilarity between two distributions in $\calP(\calX)$. 
For example, 
%as a divergence between two probability measures on $\cX$. %For two measures $\mu,\nu \in \cP(\cX)$,  $\mu\ll\nu$ means that for all measurable set $E$, if $\nu(E)=0$, then we also have $\mu(E)=0$.
%Specifically, 
the total variation distance, as a distortion, between $\mu,\nu\in\calP(\calX)$ is $\rvD_\TV(\mu,\nu)\coloneqq\int \frac{1}{2}| \frac{d  \mu}{ d  \nu}-1| \,  d \nu$. The KL divergence between $\mu,\nu~(\mu\ll\nu)$ is $\rvD_\KL(\mu,\nu)\coloneqq\int \frac{d  \mu}{ d  \nu}\log \frac{d  \mu}{ d  \nu} \,  d \nu$.
For any set $A\subseteq \calX$, we use $\delta_A$ to denote its indicator function, namely, $\delta_A(x)\coloneqq \mathbbm{1}\{x\in A\}$. Additionally, we denote $(x)_+:=\max\{x, 0\}$ and $x\wedge y:= \min\{x, y\}$.
%For a $d$-dimensional vector $(X_1,\ldots,X_d)$ and integers $1\leq i<j\leq d$, we use the shorthands $X_i^j\coloneqq (X_i,\ldots,X_j)$ and  $[j]\coloneqq \{1,2,\ldots,j\}$. 
 % ($\mu\ll\nu$)
% Given two functions $f(n)$ and $g(n)$, we say that $f(n)$ is $O(g(n))$ if there exist constants $c > 0$ and $n_0 \geq  0$ such that $f(n) \leq c g(n)$ for all $n \geq n_0$.

% \hh{Not LLM, just a time series(discrete example)/auto-regressive model/0-1 Markov Chain, asymptotic result for ergodic stationary process, consider continuous case too?, protein/RNA/DNA alphabet}
\subsection{Multi-Bit Watermark Embedding}
Let $\calX$ be a finite alphabet and let $(X_t)_{t\ge 1}$ be a discrete-time stochastic process on $\calX$, and $X^T=(X_1,\ldots,X_T)\in\calX^T$ be its first $T$ random variables, characterized by a distribution $Q_{X^T}\in\calP(\calX^T)$.
% Consider a discrete alphabet $\calX$ and a length-$T$ stochastic process $X^T$ generated according to a joint distribution $Q_{X^T}\in\mathcal{P}(\mathcal{X}^T)$, which models \textcolor{orange}{the output distribution of a large language model (LLM)}.
% To embed a multi-bit message, let $M$ be uniformly drawn from the index set $[m]\coloneqq\{1,\ldots,m\}$. 
% % \wzq{Should we define the meaning of $M=0$ here?}\hh{Only decoder outputs $\hatM=0$}
% Watermark embedding is performed at the distributional level: given the message $M$ and the data-generation
% distribution $Q_{X^T}$, the encoder $f:[m]\times \calP(\calX^T) \to \calP(\calX^T\times\calZ^T|[m])$ 
% \YM{$\calP(\calX^T\times\calZ^T|[m])$: what does this notation mean? i.e., why do you need ``conditioning on [m]'' in this notation?}
% maps $(M,Q_{X^T})$ to a joint distribution
% $P_{X^T,\zeta^T\mid M}$, thereby inducing an auxiliary stochastic process
% $\zeta^T$ on a discrete alphabet $\calZ$ and a controlled statistical
% dependence between the generated process $X^T$ and 
% $\zeta^T$.
% The auxiliary sequence $\zeta^T$ serves as side information and is assumed to be available to the decoder. 
To embed a multi-bit message, let $M$ be uniformly distributed on $[m]\coloneqq\{1,\ldots,m\}$. In addition to the generated sequence $X^T$, we will introduce an \emph{auxiliary} (decoder-observable) sequence that plays the role of side information for watermark decoding\footnote{See Appendix~\ref{sec:side-information-example} for concrete examples of how side information is used in practical watermarking schemes.}.
% \byh{Need some discussion t
Concretely, let $(\zeta_t)_{t\ge1}$ denote an auxiliary stochastic process on a finite alphabet $\calZ$, and $\zeta^T$ be the first $T$ random variables.
Watermark embedding is performed at the distributional level.
Given $M$ and the data-generation distribution $Q_{X^T}$, the encoder is a (possibly randomized) mapping
$
f:\ [m]\times \calP(\calX^T)\ \to\ \calP(\calX^T\times\calZ^T),
$
which maps $(M,Q_{X^T})$ to a conditional distribution $P_{X^T,\zeta^T\mid M}(\cdot\mid M)$.
Equivalently, for each $m\in[m]$, $f(m,Q_{X^T})$ specifies the conditional law
$P_{X^T,\zeta^T\mid M=m}\in\calP(\calX^T\times\calZ^T)$,
 inducing a controlled statistical dependence between the generated process $X^T$ and the side-information process $\zeta^T$.

% \wzq{how about this: To embed a multi-bit message, let $M$ be uniformly distributed on $[m]\coloneqq\{1,\ldots,m\}$. In addition to the generated sequence $X^T$, we will introduce an \emph{auxiliary} (decoder-observable) sequence that plays the role of side information for watermark decoding.
% Concretely, let $\calZ$ be a finite alphabet and let $\zeta^T\in\calZ^T$ denote an auxiliary stochastic process.
% Watermark embedding is performed at the distributional level.
% Given $M$ and the data-generation distribution $Q_{X^T}$, the encoder is a (possibly randomized) mapping
% $
% f:\ [m]\times \calP(\calX^T)\ \to\ \calP(\calX^T\times\calZ^T),
% $
% which maps $(M,Q_{X^T})$ to a conditional distribution $P_{X^T,\zeta^T\mid M}(\cdot\mid M)$.
% Equivalently, for each $m\in[m]$, $f(m,Q_{X^T})$ specifies the conditional law
% $P_{X^T,\zeta^T\mid M=m}\in\calP(\calX^T\times\calZ^T)$,
% thereby inducing a controlled statistical dependence between the generated process $X^T$ and the side-information process $\zeta^T$.}
% \byh{I prefer this version. Need some intro for $\zeta^T$, otherwise it is abrupt. }\wzq{Modified.}

% \wzq{This paragraph may need rephrasing as it focuses on comparing conventional watermarking with modern LLM watermarking. Do we have other examples of watermarking that operate explicitly at the distribution level?}\hh{Modified.} 
By contrast, conventional post-processing watermarking embeds the message $M$ by modifying a realized data
sequence $x^T$, typically without access to---or explicit modeling of---the underlying data-generation
distribution. In comparison, modern distributional embedding operates at the level of the generative distribution, exploiting distributional information available during generation to expand the design space of watermarking schemes while limiting unnecessary perturbations to the generated data sequence. This perspective aligns with both information-theoretic watermarking models and modern generative-model watermarking methods that modify sampling distributions rather than post-process realizations.

Notably, after watermarking,
% watermark embedding, 
the generated process $X^T$ is jointly distributed according to $P_{X^T}=\bbE_{M}[\sum_{\zeta^T\in\calZ^T}P_{X^T,\zeta^T\mid M}]$, which generally differs from the original distribution $Q_{X^T}$.  Additionally, we use the statistical divergence to measure the \emph{distortion level} of a watermarking scheme.
% To measure the \emph{distortion level} of a watermarking scheme, we use the divergence between these two distributions.
\begin{definition}[$d$-Distorted Watermarking]\label{Def: distortion}
    A watermark encoder $f$ is said to be \emph{$d$-distorted} with respect to a distortion function $\rvD$, if for any $M \in [m]$ and $Q_{X^T} \in \calP(\calX^T)$, the marginal distribution on $\calX^T$ induced by the encoder output $P_{X^T,\zeta^T|M}$ satisfies $\rvD(P_{X^T|M}, Q_{X^T})\leq d$. 
\end{definition}
Here, $\rvD$ can be any divergence that is continuous in both arguments. Common examples of such divergences include total variation distance, KL divergence, and Wasserstein distance. For $d=0$, the watermarking scheme is called \emph{distortion-free}.

% \wzq{Also feels a bit too LLM-specific here.}\hh{Modified.} 
Existing works have considered several other distortion measures, e.g., distance between  realizations \citep{860188}  or divergence between per-step conditional distributions  \citep{kuditipudi2023robust,boroujeny2024multi}. In our setup, we quantify distortion at the level of joint distributions on the stochastic process rather than individual realizations or per-step conditional distributions. This reflects the fact that watermark embedding operates by modifying the data-generation mechanism of the process rather than deterministically altering a particular realization, and realization-level distortion therefore conflates embedding effects with sampling randomness. Moreover, constraining per-step conditional distributions is unnecessarily restrictive: it reduces the design freedom and feasible region of watermarking schemes, even though generated data quality is inherently a sequence-level property. By defining distortion between joint distributions, we allow coordinated deviations across time while preserving sequence structural properties, enabling a distribution-level non-asymptotic analysis and a principled characterization of the trade-offs among detection errors, distortion, and information rate.

Moreover, to ensure \emph{strong secrecy} of the embedded watermark message, we assume that the embedded message $M$ cannot be decoded by simply observing either the watermarked sequence $X^T$ or the auxiliary sequence $\zeta^T$. 
\begin{assumption}[Strong Watermark Secrecy] \label{Ass: strong secrecy}
% [Marginal Distribution Assumption]
    % The encoder $f$ 
    Watermark embedding must ensure that both $X^T$ and $\zeta^T$ are statistically independent of the embedded message $M$. 
    % That is, $P_{X^T|M=j}=P_{X^T}$ and $P_{\zeta^T|M=j}=P_{\zeta^T}$, for all $j\in[m]$.
\end{assumption}
Under Assumption \ref{Ass: strong secrecy}, the embedded message $M$ cannot  be decoded only with $X^T$ or $\zeta^T$, i.e.,
\begin{align}
    \rvI(M;X^T)=0,~~\rvI(M;\zeta^T)=0, ~~
    \text{and}~\rvI(M;X^T,\zeta^T)=\rvI(M;\zeta^T|X^T)=\rvI(M;X^T|\zeta^T).
\end{align}
To detect if $X^T$ is watermarked, the decoder must exploit its dependency on auxiliary sequence $\zeta^T$. This corresponds to decoding with side information. 
In multi-bit watermarking, an embedded message may encode traceable information, including user/model IDs, or other sensitive metadata. Requiring strong watermark secrecy ensures such information is not extractable from the watermarked output alone, nor from the side-information alone, and can only be recovered by an authorized detector who has access to both $X^T$ and $\zeta^T$ with their correlation. Without this constraint, the payload may leak either through shifts in the observable output distribution, which leads to unauthorized decoding or easy detection, or trivially through the side-information, which makes decoding independent of the watermarking mechanism. Therefore, strong watermark secrecy formalizes a threat model and rules out degenerate constructions, allowing us to study fundamental trade-offs under distribution-preserving watermarking.

% \zs{Why we need strong watermark secrecy assumption, what is weak. what will happen if weak}

% \zs{In multi-bit watermarking, an embedded message may encode traceable information, including user/model IDs, or other sensitive metadata. Requiring strong watermark secrecy ensures such information is not extractable from the watermarked output alone, nor from the side-information alone, and can only be recovered by an authorized detector who has access to both $X^T$ and $\zeta^T$ with their correlation. Without this constraint, the payload may leak either through shifts in the observable output distribution, which leads to unauthorized decoding or easy detection, or trivially through the side-information, which makes decoding independent of the watermarking mechanism. Therefore, strong watermark secrecy formalizes a threat model and rules out degenerate constructions, allowing us to study fundamental trade-offs under distribution-preserving watermarking.}

\subsection{Statistical Detection and Message Decoding}
Within our theoretical framework, we assume that the decoder $\gamma:\calX^T\times\calZ^T \to [0\!:\!m]=\{0,1,\ldots,m\}$ receives a data-auxiliary sequence pair $(X^T,\zeta^T)$ drawn from $P_{X^T,\zeta^T\mid M}$ to recover the message $M$, i.e., $\hat{M}=\gamma(X^T,\zeta^T)$. 
If $\hat{M}=0$, the  sequence $X^T$ is decoded as unwatermarked; if $\hat{M}\in [m]$, $X^T$ is decoded as watermarked with   message $\hat{M}$. In practical algorithm design, however, we do not assume that $\zeta^T$ is directly accessible by the detector, but extracted from the data sequence. %\footnote{See Appendix~\ref{sec:side-information-example} for concrete examples of how side information is used in practical watermarking schemes.}. 
% \byh{Need some discussion to justify the use of side information. } \wzq{Maybe it'd help to add a concrete LLM watermarking example, like the green/red list as we did in our NeurIPS paper, to illustrate here.}
 This system, consisting of the encoder $f$ and  decoder $\gamma$, defines an $(m,T)$ watermarking scheme with information rate $R\coloneqq \frac{1}{T}\log m$.
According to the watermark embedding process, if the stochastic process $X^T$ is unwatermarked or naturally generated, it is independent of  $\zeta^T$; otherwise, $(X^T,\zeta^T)$ is jointly distributed according to one of the $m$ distributions $\{P_{X^T,\zeta^T|M=j}\}_{j=1}^m$. Thus, detecting and decoding the 
% watermark
message $M$ boils down to the $(m+1)$-ary hypothesis testing:
\begin{itemize}[leftmargin=*]
    \item $\rmH_0$: $X^T$ is unwatermarked, i.e., $(X^T,\zeta^T)\sim \bbP_0\coloneqq P_{X^T,\zeta^T|M=0}= Q_{X^T}\otimes P_{\zeta^T}$;
    \item $\rmH_j, \forall j\in [m]$: $X^T$ is watermarked and embedded with message $j$, i.e., $(X^T,\zeta^T)\sim \bbP_j\coloneqq P_{X^T,\zeta^T|M=j}$.
\end{itemize}
Detection performance is measured by the \emph{false-alarm error} and the \emph{$j$-th error probability}: 
    \setlength{\abovedisplayskip}{2pt}
    \setlength{\belowdisplayskip}{3pt}
    \setlength{\abovedisplayshortskip}{2pt}
    \setlength{\belowdisplayshortskip}{3pt}
\begin{align}
    \beta_0(\gamma,Q_{X^T}\otimes P_{\zeta^T})&\coloneqq\bbP_0(\gamma(X^T,\zeta^T)\ne 0 ),\\
    \beta_j(\gamma,P_{X^T,\zeta^T|M=j})&\coloneqq\bbP_j(\gamma(X^T,\zeta^T)\ne j ), \quad \forall j\in[m],
    % &=\underbrace{\bbP_j(\gamma(X^T,\zeta^T)=0)}_{\text{miss-detection error}}+\underbrace{\sum_{i\in[m]\backslash j}\bbP_j(\gamma(X^T,\zeta^T)=i)}_{\text{miss-decoding error}}, \quad \forall j\in[m].
\end{align}
where the $j$-th error probability contains both the miss-detection error probability and the miss-decoding error probability.
Let $\bar{\beta}(\gamma,(\bbP_j)_{j\in[m]})\coloneqq\frac{1}{m}\sum_{j\in[m]}\beta_j(\gamma,P_{X^T,\zeta^T|M=j})$ denote the \emph{average detection error probability}. 
There exists a natural trade-off among the false-alarm error, the average (or the max per-message) detection error probability, the distortion level and the information rate. Under Assumption \ref{Ass: strong secrecy}, we aim to characterize this trade-off by solving the following optimization problems with a feasible set defined as
% \begin{align}
%     &\min_{\gamma, (\bbP_k)_{k\in[m]}}  \; \bar{\beta}(\gamma, (\bbP_j)_{j\in[m]})  \tag{P1} \label{Eq: opt-O}
%     \\
%      &\qquad \text{s.t.}  \; \sup_{Q_{X^T}} \beta_0(\gamma,Q_{X^T}\otimes P_{\zeta^T})\leq \alpha, \quad \rvD(P_{X^T},Q_{X^T})\leq d;
% \end{align}
% and
% \begin{align}
%     &\min_{\gamma, (\bbP_k)_{k\in[m]}}  \; \max_{j\in[m]}\beta_j(\gamma, \bbP_j)  \tag{P2} \label{Eq: opt-i}
%     \\
%      &\qquad \text{s.t.}  \; \sup_{Q_{X^T}} \beta_0(\gamma,Q_{X^T}\otimes P_{\zeta^T})\leq \alpha, \quad \rvD(P_{X^T},Q_{X^T})\leq d.
% \end{align}
\newcommand{\Feas}{\mathcal{F}(\alpha,d)}
\begin{align}
\Feas \coloneqq \Big\{(\gamma,(\bbP_k)_{k\in[m]}) :\ 
\sup_{Q_{X^T}} \beta_0(\gamma,Q_{X^T}\!\otimes\! P_{\zeta^T})\le \alpha,\ 
\rvD(P_{X^T},Q_{X^T})\le d \Big\}.
\end{align}
We consider two objective criteria over the same feasible set:
\begin{align}
\tag{P1} \label{Eq: opt-O} \quad &\min_{(\gamma,(\bbP_k)_{k\in[m]})\in \Feas} \ \bar{\beta}(\gamma,(\bbP_k)_{k\in[m]}),\\
\tag{P2} \label{Eq: opt-i} \quad &\min_{(\gamma,(\bbP_k)_{k\in[m]})\in \Feas} \ \max_{j\in[m]} \beta_j(\gamma,\bbP_j).
\end{align}
Notably, we consider controlling the \emph{worst-case} false-alarm error $\sup_{Q_{X^T}} \beta_0(\gamma,Q_{X^T}\otimes P_{\zeta^T})$ since naturally generated data may vary widely across domains and sources. This includes human-written text, biological data sequences, synthetic data-generation mechanism and so on, whose underlying distributions are often heterogeneous and unknown.  

% \hh{Included more examples}

\subsection{Possible Generative Distributions}
\label{subsec:generative-distributions}

We consider in-process watermarking of stochastic processes whose generative distributions are accessible. Below, we specify the classes of discrete-alphabet processes covered by our formulation.

\paragraph{Arbitrary discrete-alphabet processes.}
In the most general setting, the generative distribution $Q_{X^T}$ is an arbitrary probability mass function (PMF) on $\calX^T$, with no assumptions on independence, stationarity, ergodicity, or memory. This setting encompasses general discrete stochastic processes, including non-stationary sources and distributions induced by complex generative mechanisms.

\paragraph{Stationary ergodic processes.}
% As a special case, we consider stationary ergodic stochastic processes. Let $\{X_t\}_{t\ge 1}$ be a discrete-time stochastic process taking values in $\calX$. The process is said to be \emph{stationary} if for all $k\ge 1$, all $t\ge 1$, and all $x^k\in\calX^k$, $\Pr(X^k = x^k) = \Pr(X_t^{t+k-1} = x^k)$.
% The process is said to be \emph{ergodic} if every shift-invariant event has probability either zero or one. Equivalently, time averages of integrable functions converge almost surely to their ensemble averages.
As a structured special case, we consider stationary ergodic processes. A discrete-time process $(X_t)_{t\ge1}$ on $\calX$ is \emph{stationary} if its finite-dimensional distributions are invariant under time shifts, and \emph{ergodic} if every shift-invariant event has probability zero or one.

For a stationary process $Q$, the \emph{entropy rate} is defined as 
\begin{equation}
    \rvH_Q(\calX) \;\coloneqq\; \lim_{T\to\infty} \frac{1}{T} \rvH(X^T),
\end{equation}
%\YM{$\rvH(\calX)$ is a strange notation, since ${\cal X}$ is not the random process considered. How about $\rvH(\{X_t\})$, or complying with your earlier notation $\rvH\left((X_t)_{t\ge 1}\right)$}\hh{I copy this notation from Cover and Thomas.} \YM{Indeed, that is his notation. then be it.}
provided the limit exists. For stationary ergodic processes with a finite alphabet, this limit exists and is equal to $\rvH_Q(\calX) = \lim_{T\to\infty} \rvH(X_T \mid X^{T-1})$.
% \[
% \rvH(\calX) = \lim_{T\to\infty} \rvH(X_T \mid X^{T-1}).
% \]

% Throughout the paper, unless otherwise stated, the generative distribution $Q_{X^T}$ may belong to either of the above classes.

\section{Fundamental Trade-Offs for 
% Arbitrary 
Stochastic Processes with Finite Length}\label{sec: finite length}
\setcounter{theorem}{0}

% \zs{highlight generalized case, a special case: deterministic decoding. 4.1 randomized (converse+ constructive proof), (connection between 4.1 and 4.2) 4.2 deterministic (achievability)}

% \subsection{Finite-Length Analysis for Arbitrary Stochastic Processes}

% \subsubsection{Results under Strong Secrecy Assumption \ref{Ass: strong secrecy}}
In this section, we characterize the fundamental trade-offs among the detection error, false-alarm error, distortion and information rate through the optimal objective values of \eqref{Eq: opt-O} and \eqref{Eq: opt-i}. We begin by deriving the optimal detection error probability by allowing randomized decoders, established via matching converse and achievability results. This characterization provides a fundamental lower bound that applies to all deterministic decoders defined in our problem formulation. We then present achievability results for deterministic decoders, which approach the lower bound up to a small (and in some cases zero) residual error.

Specifically, we temporarily relax the decoding rule and allow \emph{soft (randomized) decoding}, i.e., randomized decoders $\gamma(\cdot|x^T,\zeta^T)\in\calP([0:m])$. Under this decoding strategy, the corresponding false-alarm error and $j$-th error probability are defined as
\begin{align}
    \beta_0(\gamma,Q_{X^T}\otimes P_{\zeta^T})\coloneqq \bbE_{\bbP_0}[1-\gamma(0|X^T,\zeta^T)], \quad \beta_j(\gamma, P_{X^T,\zeta^T|M=j})\coloneqq \bbE_{\bbP_j}[1-\gamma(j|X^T,\zeta^T)].
\end{align}
When $\gamma$ is restricted to output one-hot vectors (i.e., $\gamma(j\mid x^T,\zeta^T)\in\{0,1\}$ for all $j$), this formulation reduces to the deterministic decoder defined in the problem formulation.

We emphasize that this relaxation to randomized decoding is solely for the purpose of establishing tight fundamental limits via matching converse and achievability arguments. In practical watermarking systems, the decoder is typically deterministic. As we will show later, the performance characterized under randomized decoding serves as a fundamental lower bound, which can be approached arbitrarily closely
(and in some cases exactly achieved) by deterministic schemes.
\begin{theorem}[Optimal Detection Error Probability]\label{thm: opt ave beta strong}
   Under Assumption \ref{Ass: strong secrecy} and allowing randomized decoders, the minimum average detection error probability $\bar{\beta}^*(Q_{X^T},\alpha,d,m)$ obtained from \eqref{Eq: opt-O} and the min-max  per-message error detection probability $\beta_{\max}^*(Q_{X^T},\alpha,d,m)$ obtained from \eqref{Eq: opt-i} coincide 
   and admit the same value
    \begin{align}
        % \LB(Q_{X^T},\alpha,d,m)
        \bar{\beta}^*(Q_{X^T},\alpha,d,m)=\beta_{\max}^*(Q_{X^T},\alpha,d,m)
        % \coloneqq
        =\min_{P_{X^T}:\rvD(P_{X^T},Q_{X^T})\leq d}\sum_{x^T\in\calX^T}\left(P_{X^T}(x^T)-\tfrac{\alpha}{m}\right)_+.\label{eq: dec error opt}
    \end{align}
    % \YM{In above, shouldn't ``$\coloneqq$'' be changed to  ``$=$''?}
    Moreover, if the distortion measure is chosen as the total variation distance $\rvD_\TV(P_{X^T},Q_{X^T})$,  the optimal value admits the closed-form expression
    \begin{equation}
        \bigg(\sum_{x^T}\left(Q_{X^T}(x^T)- \tfrac{\alpha}{m}\right)_+ -\Big(d \wedge \sum_{x^T}\big( \tfrac{\alpha}{m}-Q_{X^T}(x^T)\big)_+ \Big)\bigg)_+.
    \end{equation}
\end{theorem}
Theorem \ref{thm: opt ave beta strong} characterizes a fundamental trade-off among the average (or the max per-message) detection error probability, false-alarm error, distortion, and information rate. Specifically, for a fixed blocklength $T$, increasing the false-alarm error $\alpha$ or the distortion level $d$, or decreasing the information rate $R=\log m / T$ (for a fixed $T$), allows for a smaller detection error probability. This trade-off offers a quantitative benchmark for evaluating watermarking methods in stochastic processes, e.g., watermarking for generative models, and identifies an impossibility regime determined by the underlying model and secrecy constraints.
We will further demonstrate the trade-off among the four facets graphically in Section \ref{Sec: simulations}.

% Moreover, Theorem \ref{thm: opt ave beta strong} reveals a  consequence of the secrecy constraint. Since Assumption \ref{Ass: strong secrecy} enforces that the marginal distribution of the observable sequence $X^T$ is identical under all messages, the decoder cannot statistically distinguish among different watermark messages based on the observation alone. As a result, the decoder is forced to treat all messages symmetrically, which implies that the optimal average detection error probability and the optimal maximum per-message failure probability coincide. 
% The expression of the optimal failure probability also highlights that the false-alarm budget $\alpha$ is equally allocated under each message. Therefore, increasing the number of messages uniformly tightens the effective worst-case false-alarm control for each message, which leads to a higher detection error probability. When $m=1$, the result reverts back to that in the zero-bit watermarking setting \citep{he2025theoretically}.
Moreover, Theorem~\ref{thm: opt ave beta strong} reveals an consequence of the secrecy constraint. Under Assumption~\ref{Ass: strong secrecy}, the observable distribution of $X^T$ is identical across messages, preventing the decoder from statistically distinguishing among them based on the observation alone. Consequently, all messages must be treated symmetrically, and the optimal average detection error coincides with the optimal maximum per-message error. The resulting expression further shows that the false-alarm budget $\alpha$ is evenly allocated across messages, so increasing the message set size tightens the effective false-alarm constraint per message and increases the detection error probability. Note that the special case $m=1$ recovers the zero-bit watermarking result of~\citet{he2025theoretically}.

The proof of Theorem \ref{thm: opt ave beta strong} consists of a converse and an achievability. In the converse (proved in  Appendix \ref{App: coverse of thm: opt ave beta strong}), we establish a fundamental lower bound for the optimal objective values of~\eqref{Eq: opt-O} and~\eqref{Eq: opt-i}, which holds for both randomized and deterministic decoders. In the achievability, we show that, for a feasible regime of rate, there exist watermarking schemes with randomized decoders attaining this lower bound. When the decoder is deterministic, we propose an explicit scheme that approaches the lower bound up to a small (and in some cases zero) residual error.

\paragraph{Converse (Sketch) for Theorem \ref{thm: opt ave beta strong}.} The converse establishes a lower bound on the average and maximum per-message detection error probabilities by first fixing an arbitrary randomized decoder and exploiting the constraint that the worst-case false-alarm error is bounded above by $\alpha$. We then show that this lower bound is optimized when the false-alarm budget is evenly allocated across messages and the watermarked source distribution is optimized, yielding a bound that depends only on $(Q_{X^T}, d, \alpha, m)$ and is independent of the 
% specific
decoder or watermarking joint distribution. As a result, the bound holds uniformly over the entire feasible region. See Appendix~\ref{App: coverse of thm: opt ave beta strong}  for a complete proof.

\subsection{Achievability for Theorem \ref{thm: opt ave beta strong}}
% In the converse,  the optimal detection error is lower bounded by the RHS of \eqref{eq: dec error opt} for any $m$. In this section, we prove the existence of watermarking schemes with randomized decoders that achieve this lower bound, and further provide an explicit scheme with a deterministic decoder that attains the bound up to a small residual error $\varepsilon_{\text{res}}$ for $m$ in a feasible regime.
% The proposed embedding–detection scheme is intentionally simple: it serves as a constructive realization of the theoretical benchmark and demonstrates the achievability of the derived bounds, rather than aiming for state-of-the-art empirical performance. Before stating the following theorem, we let $P^*_{X^T}$ be an optimizer of~\eqref{eq: dec error opt}: $P^*_{X^T}\coloneqq \argmin_{P_{X^T}: \rvD(P_{X^T},Q_{X^T})\leq d} \sum_{x^T} (P_{X^T}(x^T)-\tfrac{\alpha}{m})_+$,
% \begin{equation}
%     P^*_{X^T}\coloneqq \argmin_{P_{X^T}: \rvD(P_{X^T},Q_{X^T})\leq d} \sum_{x^T} \bigg(P_{X^T}(x^T)-\frac{\alpha}{m}\bigg)_+,
% \end{equation}
% and define $\calS_{\alpha/m}\coloneqq\{x^T: P^*_{X^T}(x^T)\ge \tfrac{\alpha}{m}\}$.

% \wzq{I attempt to break up the original Theorem into two parts:

In the converse,  the optimal detection error is lower bounded by the RHS of \eqref{eq: dec error opt} for any $m$. In this section, we construct the watermarking schemes with randomized decoders that achieve this lower bound, and further provide an explicit scheme with a deterministic decoder that attains the bound up to a small residual error $\varepsilon_{\mathrm{res}}$ for $m$ in a feasible regime.
The proposed embedding--detection scheme is intentionally simple: it serves as a constructive realization of the theoretical benchmark and demonstrates the achievability of the derived bounds, rather than aiming for state-of-the-art empirical performance. Before stating the main results, we let $P^*_{X^T}$ be an optimizer of~\eqref{eq: dec error opt}: $P^*_{X^T}\coloneqq \argmin_{P_{X^T}: \rvD(P_{X^T},Q_{X^T})\leq d} \sum_{x^T} (P_{X^T}(x^T)-\tfrac{\alpha}{m})_+$,
% \begin{equation}
%     P^*_{X^T}\coloneqq \argmin_{P_{X^T}: \rvD(P_{X^T},Q_{X^T})\leq d} \sum_{x^T} \bigg(P_{X^T}(x^T)-\frac{\alpha}{m}\bigg)_+,
% \end{equation}
and define the ``heavy'' set $\calS_{\alpha/m}\coloneqq\{x^T: P^*_{X^T}(x^T)\ge \tfrac{\alpha}{m}\}$.

\subsubsection{Randomized Decoding}

\begin{theorem}[Optimal Watermark Embedding and Detection: Randomized Decoder]
\label{thm: ach rand scheme}
Assume Assumption~\ref{Ass: strong secrecy}. 
Fix $Q_{X^T}$, false-alarm constraint $\alpha$, distortion level $d$, and message set size
$m\leq m^*\coloneqq \max\{m\in[|\calX|^T]:\bar{\beta}^*(Q_{X^T},\alpha,d,m)\leq 1-\alpha\}$.
Under randomized decoding, there exist a watermark embedding distribution
$P^*_{\zeta^T|X^T,M}$ and a randomized decoder
$\gamma(\cdot | x^T,\zeta^T)\in\calP([0:m])$ that satisfy the distortion and false-alarm constraints, and achieve the optimal value in~\eqref{eq: dec error opt}.
\end{theorem}
% Theorem \ref{thm: ach rand scheme} shows that to achieve the detection-error lower bound uniformly for all generative distributions $Q_{X^T}$, it is therefore necessary to allow \textbf{randomized constructions}, specifically, a randomized decoder $\gamma$ and a stochastic grouping kernel $\phi$. 
Theorem~\ref{thm: ach rand scheme} identifies the information-theoretic limit when randomized decoding is permitted: the achieved detection performance matches the converse benchmark~\eqref{eq: dec error opt} uniformly over all $Q_{X^T}$. 
From a constructive viewpoint, randomization effectively enables ``fractional'' allocation of probability mass at the threshold $\alpha/m$, which is crucial for matching the converse exactly.
This theorem therefore serves as the baseline against which we evaluate deterministic decoding.

In practice, however, watermark decoders are typically deterministic.
One of our main technical contributions is an explicit deterministic construction that approximates the randomized optimum. 
The key idea is to first \emph{group} source sequences $x^T$ into a finite number of bins and then carry out watermark embedding and decoding at the \emph{group level}.
Intuitively, this grouping step creates at least $m$ ``heavy'' groups with mass $\ge\alpha/m$, which can reliably support the $m$ messages; the side information $\zeta^T$ is then used to assign messages within heavy groups, while the decoder outputs $0$ whenever the pair $(x^T,\zeta^T)$ falls outside the designed decodable region.

\paragraph{Construction (grouping $\to$ embedding $\to$ decoding).}
Let $K\ge m$ and let $\phi:\calX^T\to[K]$ be a deterministic grouping function, inducing a group distribution $P_G$ via $P_G(k)=P^*_{X^T}(\phi(X^T)=k)$.
Define $\calG_{\alpha/m}\coloneqq\{k\in[K]:P_G(k)\ge \alpha/m\}$.

\smallskip
\noindent\textbf{Step 1 (Grouping).}\quad
Choose $\phi^\star$ as a solution to
\begin{align}
\min_{\phi:\calX^T\to[K]} \quad
\sum_{k=1}^K \Bigl(P_G(k)-\tfrac{\alpha}{m}\Bigr)_+
\quad
\text{s.t.}\quad
|\{k:P_G(k)\ge\tfrac{\alpha}{m}\}|\ge m,
\label{eq: opt phi thm}
\end{align}
with the additional constraint that every $x^T\in\calS_{\alpha/m}$ is assigned to a singleton group.

\smallskip
\noindent\textbf{Step 2 (Embedding).}\quad
Let $\calZ^T=\calZ^T_{\ge}\cup\calZ^T_{<}\cup\{\widetilde{\zeta^T}\}$ be a finite alphabet,
where $|\calZ^T_{\ge}|=|\calG_{\alpha/m}|$, $|\calZ^T_{<}|=K-|\calG_{\alpha/m}|$ and $\widetilde{\zeta^T}$ is an additional redundant state.
For each $j\in[m]$, define
\[
P^*_{\zeta^T|X^T,M}(\zeta^T|x^T,j)
=
P^*_{\zeta^T|G,M}(\zeta^T|\phi^\star(x^T),j),
\]
where for all $k\in[K]$,
\begin{align}
P^*_{\zeta^T|G,M}(\zeta^T|k,j)
&=\Bigl(1\wedge \tfrac{\alpha}{mP_G(k)}\Bigr)\mathbbm{1}\{h(k,\zeta^T)=j\}
+ \mathbbm{1}\{\zeta^T\in\calZ^T_<,\, g(\zeta^T)=k\} \notag\\
&\quad + \Bigl(1-\tfrac{\alpha}{mP_G(k)}\Bigr)_+ \mathbbm{1}\{\zeta^T=\widetilde{\zeta^T}\},
\label{eq: opt embed thm}
\end{align}
with $h:\calG_{\alpha/m}\times\calZ^T_{\ge}\to[|\calG_{\alpha/m}|]$ bijective in either argument when the other is fixed, and
$g:\calZ^T_{<}\to[K]\setminus\calG_{\alpha/m}$ bijective.

\smallskip
\noindent\textbf{Step 3 (Decoding).}\quad
Define the deterministic decoder $\gamma^\star:\calX^T\times\calZ^T\to\{0,1,\dots,m\}$ by
\begin{equation}
\gamma^\star(x^T,\zeta^T)=
\begin{cases}
j, & \text{if } \zeta^T\in\calZ^T_{\ge}\text{ and } h(\phi^\star(x^T),\zeta^T)=j\le m,\\
0, & \text{otherwise}.
\end{cases}
\label{eq: opt dec thm}
\end{equation}

\subsubsection{Deterministic Decoding}
We are now in a position to present our achievability result under deterministic decoding.

\begin{theorem}[Deterministic Decoding: Achievability and Residual Error]
\label{thm: ach det scheme main}
Under Assumption~\ref{Ass: strong secrecy} and fix $Q_{X^T}$, $\alpha$, and $d$.
For any $m\in\calM\coloneqq \Bigl\{m\in [|\calX|^T]:
\sum_{x^T\notin \calS_{\alpha/m}}P^*_{X^T}(x^T)\geq \bigl(1-|\calS_{\alpha/m}|/m\bigr)\alpha\Bigr\}$,
the above construction produces a deterministic decoder $\gamma^\star$ and an embedding distribution $P^*_{\zeta^T|X^T,M}$ that satisfy the distortion and false-alarm constraints. 
Moreover, the resulting detection error matches the randomized converse benchmark up to a residual term:
\begin{equation}
\beta_{\mathrm{det}}(Q_{X^T},\alpha,d,m)
\le \bar{\beta}^*(Q_{X^T},\alpha,d,m) + \varepsilon_{\mathrm{res}}(Q_{X^T},\alpha,d,m),
\label{eq: det residual}
\end{equation}
where $\varepsilon_{\mathrm{res}}$ (cf. \eqref{Eq: res error} in Appendix \ref{App: determ dec proof}) is the residual error incurred by deterministic grouping on the discrete alphabet $\calX^T$. In particular, $\varepsilon_{\mathrm{res}}=0$ whenever the grouping constraint admits exact mass balancing.
\end{theorem}
Theorem~\ref{thm: ach det scheme main} formalizes the price of determinism. Under strong secrecy, the side information $\zeta^T$ cannot carry message information by itself and effectively acts as shared randomness; hence reliable multi-bit decoding must be supported by the observed process $X^T$.
Randomized decoding can match the converse benchmark exactly, while deterministic decoding approximates it via a concrete grouping-based design.
In general, a nonzero residual error $\varepsilon_{\mathrm{res}}$ is unavoidable: deterministic grouping partitions the finite set $\calX^T$ and therefore cannot, in general, realize an exact ``fractional'' allocation of probability mass needed to match the randomized optimum for arbitrary $Q_{X^T}$.

We restrict attention to $m\le |\calX|^T$ and do not consider $m>|\calX|^T$.
Indeed, since $X^T$ takes values in the finite alphabet $\calX^T$, the decoder observes at most $|\calX|^T$ distinct realizations.
Under deterministic decoding and strong secrecy ($P_{\zeta^T|M=j}=P_{\zeta^T}$ for all $j$), the auxiliary randomness $\zeta^T$ cannot convey message information by itself; thus message distinguishability must come from $X^T$, which fundamentally limits reliable recovery to $m\le |\calX|^T$ (equivalently, $R=\frac{\log m}{T}\le \log|\calX|$).
If $m>|\calX|^T$, distinct messages are necessarily indistinguishable from the observation under any deterministic scheme unless one relaxes the decoding criterion or enlarges the observation model.

Finally, we note that the feasible regime of $m$ also depends on the optimal source distribution $P^*_{X^T}$ and the false-alarm constraint $\alpha$.
In particular, when $P^*_{X^T}$ is uniform over $\calX^T$, it is possible to support the maximal message set size $m=|\calX|^T$.

\section{Fundamental Trade-Offs for Stationary Ergodic Processes with Infinite Length}\label{sec: asymptotics}
In this section, we present the asymptotic analysis for stationary ergodic processes by letting the blocklength $T \to \infty$. Specifically, let $P$ and $Q$ be stationary ergodic measures on $\calX^\bbN$.
The resulting characterization reveals the fundamental
performance limits achievable under stationarity and ergodicity, and serves as a benchmark for
understanding the finite-length trade-offs developed earlier.
% established in the previous sections.
We will
explicitly compare the asymptotic and finite-length results at the end of this section.
Define the relative entropy rate between two stationary ergodic processes as $\overline \rvD_\KL(P\|Q)\coloneqq \lim_{T\to\infty}\frac{1}{T}\rvD_\KL(P_{X^T}\|Q_{X^T})$.

\begin{theorem}[Asymptotically Optimality for Stationary Ergodic Processes]\label{thm: asymp opt rate and dec}
    Assume Assumption~\ref{Ass: strong secrecy}. Fix $Q$ be a stationary ergodic process (SEP) on the finite
alphabet $\calX$, false-alarm constraint $\alpha$, distortion level $d$. For any SEP $P$ satisfying 
$\rvD(P_{X^T},Q_{X^T})\le dT$ for all $T$,
and any $(dT)$-distorted $(m,T)$ watermarking scheme with message set $[m]$ and any associated
tuple $(\bbP_k)_{k\in[m]}$, if the average detection error
$\bar{\beta}(\gamma,(\bbP_k)_{k\in[m]}) \to 0$ as $T\to\infty$, then the achievable information
rate is upper bounded by the maximum entropy rate:
   \begin{align}
       R\leq \max_{P:\overline\rvD_\KL(P \| Q)\leq d}\rvH_P(\calX).
   \end{align}
   % where the last bound holds for all $d$-distorted $(m,T)$ watermarking schemes for the LLM $Q_{X^T}$.

When $d=0$, for any $m\leq \alpha\exp(T\rvH_Q(\calX))$, there exist an asymptotically optimal deterministic decoder $\gamma^*$, and an optimal distortion-free watermarking scheme that achieve the detection performance:  
as $T\to\infty$, for all $j\in[m]$,
\begin{align}
    \beta_j(\gamma^*, \bbP_j^*)&\leq \exp(-\Omega(T^{\frac{1}{2}}))\to 0, \label{eq: beta_j asymp} 
    \quad \sup_{Q_{X^T}}\beta_0(\gamma^*, \bbP_0^{*})\leq\alpha+\exp(-\Omega(T^{\frac{1}{2}}))\to \alpha.
\end{align}
\end{theorem}
Theorem \ref{thm: asymp opt rate and dec} presents the asymptotically optimal information rate for watermarking on a stationary ergodic process, which is equal to its entropy rate when $d=0$. This maximum information rate coincides with the steganography capacity characterized in classical works  without adversary and noise \citep{4418489,harmsen2009capacity}. As the distortion increases, a $(dT)$-distorted watermarking can trade off text quality to achieve a higher rate. 

% If we choose the distortion function $\rvD$ as KL divergence $\rvD_\KL$, we have the following result regarding optimizing the stationary ergodic process $P_{X^T}$ over the feasible set $\rvD_\KL(P_{X^T}\|Q_{X^T})\leq d$ for all $T\in\bbN$.
% \begin{lemma}\label{Lem: distortion asymp}
%     Let $P$ and $Q$ be stationary ergodic process measures on $\calX^\bbN$ such that $P_{X^T}\ll Q_{X^T}$ for all $T$. If $\rvD_\KL(P_{X^T}\|Q_{X^T})\leq d$ for all $T\in\bbN$, then $P=Q$. In particular, their entropy rate coincides $\rvH_P(\calX)=\rvH_Q(\calX)$.
% \end{lemma}
% The proof of Lemma \ref{Lem: distortion asymp} is provided in Appendix \ref{App: pf of Lem: distortion asymp}. We optimize over stationary ergodic process measures $P$ satisfying a KL distortion
% constraint $\rvD_\KL(P_{X^T}\|Q_{X^T})\le d$ for all $T\in\bbN$. This implies the relative entropy rate
% $\overline \rvD_\KL(P\|Q)\triangleq \limsup_{T\to\infty}\frac{1}{T}\rvD_\KL(P_{X^T}\|Q_{X^T})$ equals zero, hence
% $P=Q$ for finite-alphabet stationary processes. Therefore the feasible set contains only $Q$.

\paragraph{Proof Sketch for Theorem \ref{thm: asymp opt rate and dec}.} Here we provide main ideas of the converse and achievability proofs.
\begin{itemize}[itemsep=0pt,topsep=0pt,leftmargin=12pt]
    \item \textbf{Converse:} We apply Fano's inequality to obtain the maximum information rate that a $dT$-distorted $(m,T)$ watermarking scheme can achieve on stochastic ergodic processes, with vanishing average detection error $\bar{\beta}(\gamma,(\bbP_k)_{k\in[m]})$. Specifically, we leverage the strong secrecy Assumption \ref{Ass: strong secrecy}, where $\rvH(M)=\rvH(M|\zeta^T)$. The details are in Appendix \ref{App: pf of best rate}.

    \item \textbf{Achievability:} We identify the asymptotically optimal watermarking scheme that achieves vanishing $j$-th error probabilities and the maximum information rate, while satisfying the worst-case false-alarm error and distortion constraint. In particular, by applying the method of types and asymptotic equipartition property in information theory, the asymptotically optimal decoder deterministically maps a typical sequence $x^T$ to a typical sequence $\zeta^T$ uniquely under different messages $M$. The corresponding optimal joint distribution $P_{X,\zeta|X,M}^{*T}$ assigns probability $1$ to such pair of sequences $(x^T,\zeta^T)$, making sure that the detection accuracy is high. The details are  in  Appendix \ref{App: pf of thm: asymp opt rate and dec}.
\end{itemize}

\begin{remark}[Comparison with Finite-Sample Analysis]
    Consider Theorem~\ref{thm: opt ave beta strong} under zero distortion ($d=0$). For certain source distributions $Q_{X^T}$, the maximum achievable message set size (in bits) is $\log m^* = \rvH(Q_{X^T})$, which coincides with the asymptotic entropy-rate characterization. For example,  when $Q_{X^T}$ is uniform over $\calX^T$, we have $\log m^*=\log|\calX|^T$. Under certain under appropriate scaling of the false-alarm constraint, the finite-sample optimal detection error characterized in Theorem~\ref{thm: opt ave beta strong} also aligns with the asymptotic result.   Under the same example,  the optimal detection error probability in \eqref{eq: dec error opt} reduces to
        $\sum_{x^T}\left(\frac{1}{|\calX|^T}-\frac{\alpha}{|\calX|^T}\right)_+=1-\alpha.$
    If the false-alarm constraint $\alpha=1-\exp(-\Omega(T^{\frac{1}{2}}))$ increases to $1$ exponentially fast with $T$, then the finite-sample optimal detection error exhibits the same decay behavior as the asymptotic detection error in Theorem~\ref{thm: asymp opt rate and dec}. %In this sense, the finite-sample characterization recovers the asymptotic performance limits under appropriate scaling of the false-alarm constraint.
\end{remark}

\section{Numerical Results}\label{Sec: simulations}
In this section, we numerically compute and visualize the trade-offs characterized in Theorem~\ref{thm: opt ave beta strong}. We also illustrate the gap between the deterministic-decoder achievability bound and the converse, and compare the finite-sample information rate to the asymptotically optimal rate for stationary ergodic stochastic processes, i.e., the entropy rate. For computational tractability, we consider a finite-state stationary ergodic Markov model.

\begin{figure}[t]
    \centering
    \subfloat[]{\label{fig:beta_vs_r}\includegraphics[width=0.33\linewidth]{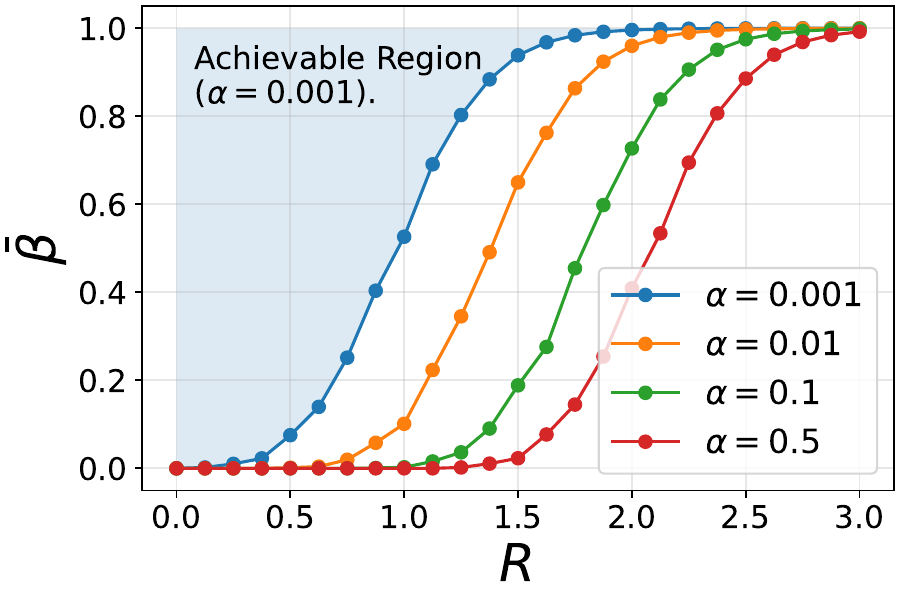}}
    \subfloat[]{\label{fig:beta_vs_alpha}\includegraphics[width=0.33\linewidth]{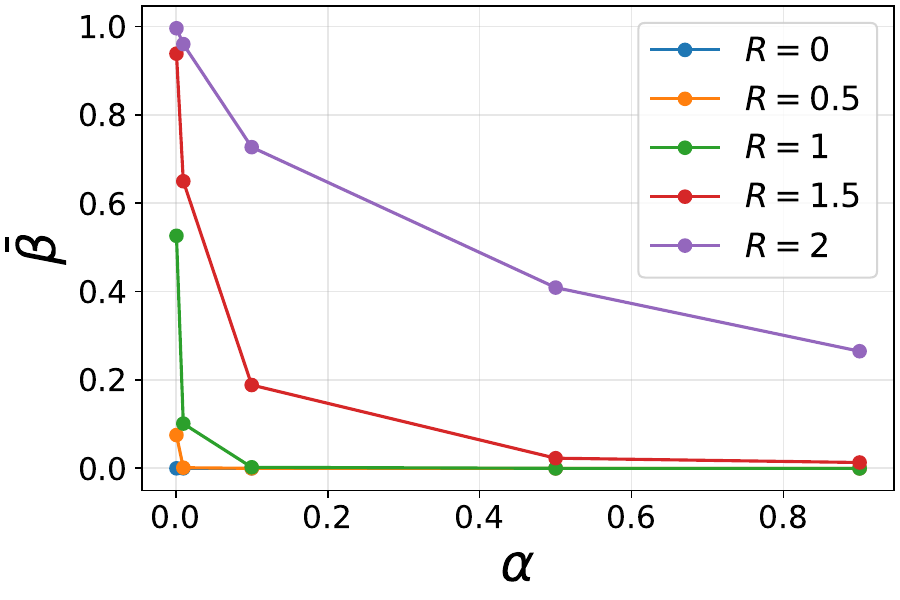}}
    \subfloat[]{\label{fig:beta_vs_r_res}\includegraphics[width=0.33\linewidth]{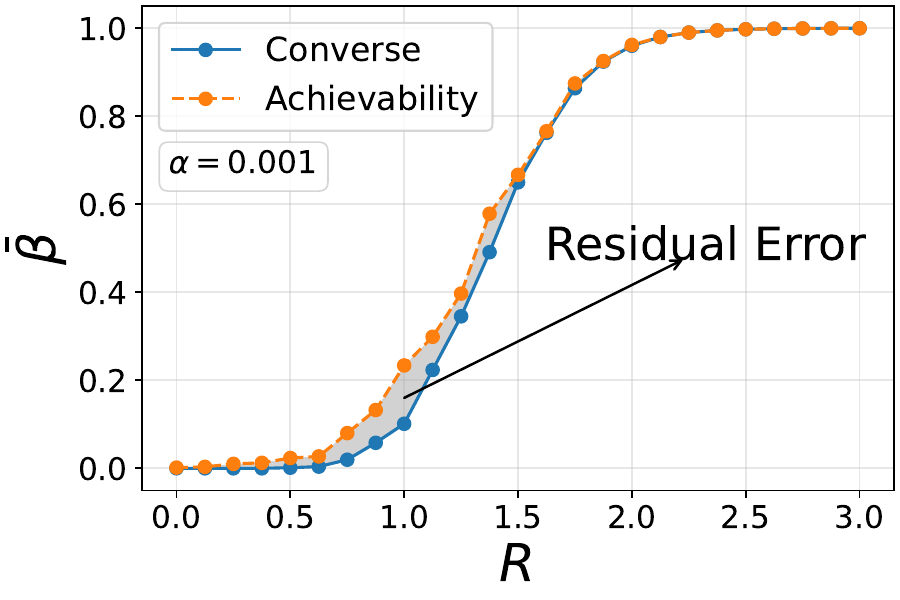}}
    
    \vspace{-0.8em}
    
    \subfloat[]{\label{fig:beta_vs_alpha_res}\includegraphics[width=0.33\linewidth]{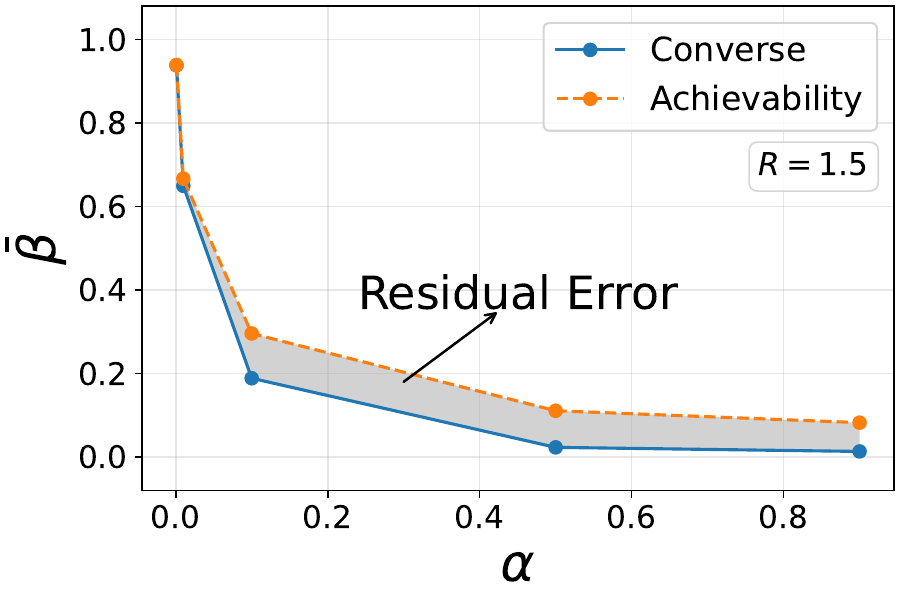}}
    \subfloat[]{\label{fig:r_vs_t}\includegraphics[width=0.33\linewidth]{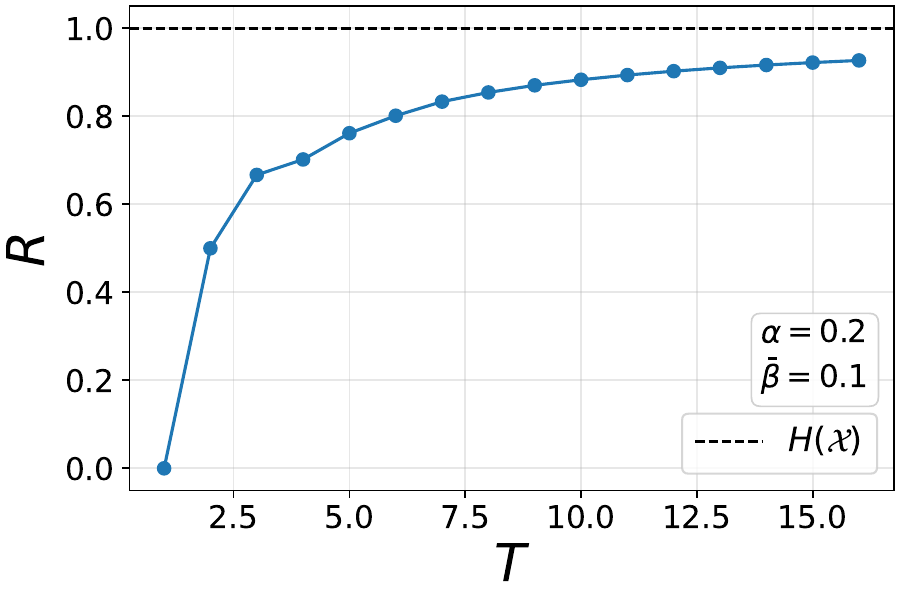}}
    \subfloat[]{\label{fig:beta_vs_distortion}\includegraphics[width=0.33\linewidth]{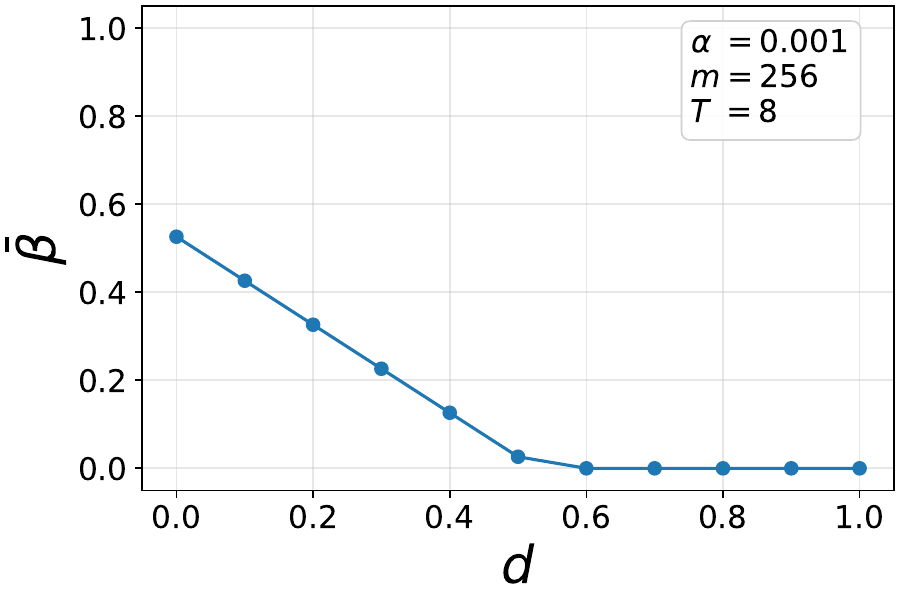}}
    % \vspace{-0.7em}
    \caption{Numerical results on stationary ergodic Markov chain. }
    % \vspace{-1.5em}
\end{figure}

We first present the trade-offs across three facets $(\alpha,\bar{\beta}, R)$ under zero distortion ($d=0$) for a fixed-length process. Figure \ref{fig:beta_vs_r} presents the \textbf{rate-detection} trade-off under a fixed false-alarm error $\alpha$. Clearly, as the rate $R$ increases, the average detection error probability increases as well. The region above each curve is the achievable region for any watermarking scheme within our framework. As $\alpha$ increases, the achievable region enlarges. 
Figure \ref{fig:beta_vs_alpha}  presents the \textbf{false alarm-detection} trade-off under a fixed information rate $R$. The average detection error decreases as the false-alarm error increases. Similarly, the region above each curve is the achievable region. When the rate $R$ is small enough, almost the entire region is achievable. 

Figures~\ref{fig:beta_vs_r_res} and \ref{fig:beta_vs_alpha_res} illustrate the \textbf{gap} between the converse lower bound on the average detection error and the performance achieved by the reference watermarking scheme with a deterministic decoder 
(cf.~Theorem~\ref{thm: ach det scheme main}),
% (cf.~Theorem \ref{thm: ach det scheme}), 
as quantified by the residual error. We observe that this residual error is small, indicating that the proposed reference scheme---despite its simplicity---achieves performance close to the fundamental limit.

Figure \ref{fig:r_vs_t} shows the \textbf{length-rate} trade-off and that the finite-length achievable information rate approaches the fundamental limit, the entropy rate, as the sequence length $T$ increases under certain generative distributions. Figure \ref{fig:beta_vs_distortion} further shows the \textbf{distortion-detection} trade-off when considering total variation distortion. Clearly, by allowing higher distortion level, we have more budget to optimize the watermark embedding distribution and enhance watermark detection performance.

These numerical results provide a concrete instantiation of our theoretical framework, illustrating the fundamental information-theoretic trade-offs governing multi-bit watermark embedding in stochastic processes generated from a prescribed distribution. Together with our converse and achievability results, they validate the tightness of the derived bounds and clarify the impact of false-alarm control, distortion, and information rate in finite and asymptotic regimes. Beyond serving as a proof of concept, our analysis is intended to guide the design of future practical watermarking algorithms that operate at the distributional level and approach fundamental performance limits.

\clearpage

% Acknowledgments---Will not appear in anonymized version
% \acks{We thank a bunch of people and funding agency.}

\bibliography{sample.bib, IT_references.bib}
\bibliographystyle{unsrtnat}

\appendix

\clearpage
% \crefalias{section}{appendix} % uncomment if you are using cleveref

\section{Related Works}\label{Sec: related works}
Watermarking techniques have been widely explored across various stochastic processes, including text \citep{liu2024survey, liu2025position,zhao2024sok}, image \citep{ding2024erasing, liu2024image, hui2025autoregressive}, and protein sequence generation \citep{chen2025enhancing} to support traceability and detection of synthetic content. In general, existing approaches fall into two main categories: zero-bit watermarking for mere detection and multi-bit watermarking for conveying auxiliary messages. 

\paragraph{Zero-bit Watermark.} A zero-bit watermark is a type of watermarking scheme that only indicates whether a watermark is present or not, without encoding any additional information. One line of work embeds a watermark into a sequence during the decoding process, for example, by manipulating the decoding strategy \citep{kirchenbauer2023watermark, zhao2023provable, liu2024adaptive,liu2025dataset, dathathri2024scalable,kuditipudi2023robust}, applying post-training watermarking methods \citep{xu2024learning,zhao2023protecting}, or setting a watermarking instruction \citep{liu2025context,dasgupta2024watermarking}. Another line of work post-processes a generated sequence to embed a watermark \citep{an2025defending, an2025reinforcement, chang2024postmark,zhao2025tab}. Moreover, some works provide a theoretical standpoint instead of a heuristic design of the watermark \citep{cai2025optimal,li2025robust,he2025empirical,li2025optimal, li2024statistical, tsurheavywater, huang2023towards, huang2025watermarking}. Specifically, \citet{christ2024pseudorandom} applies pseudorandom error-correcting codes for watermarking to make watermarked text computationally indistinguishable from unwatermarked ones to efficient adversaries. \citet{francati2025coding} establishes optimal robustness guarantees for watermarking schemes based on pseudorandom error-correcting codes. \citet{he2025theoretically} proposes a theoretical framework that jointly optimizes the watermarking scheme and detector, showing that watermarks should be adaptive to the LLM’s output distribution to achieve the best detectability–distortion trade-off. \citet{li2024statistical} casts LLM watermark detection as hypothesis testing with pivotal statistics, providing provable Type-I error control and minimax-optimal detection under distributional uncertainty.

% The systematic use of pseudorandom error-correcting codes for watermarking was initiated by \citep{christ2024pseudorandom}. Their work introduced a framework in which watermarked texts are computationally indistinguishable from unwatermarked ones to efficient adversaries. Subsequent work establishes optimal robustness guarantees for watermarking schemes based on pseudorandom error-correcting codes \citep{francati2025coding, he2025empirical}.

% Specifically, \citet{kirchenbauer2023watermark} divides the vocabulary of LLM into green and red lists and slightly prompts the usage of data in the green list in the sampling process. \citet{gumbel2023} employs the pseudo-random sampling strategy to sample the next data of LLMs. 

\noindent\paragraph{Multi-bit Watermark.} 
%A multi-bit watermark embeds multiple bits of information into the carrier, enabling not only watermark detection but also the transmission of extra data such as an identifier or message 
Multi-bit watermarking embeds multiple bits to enable both detection and message transmission \citep{he2025distributional, li2024identifying}. Compared to the extensive literature on zero-bit watermark, research on multi-bit watermark remains sparse, with most existing methods relying on heuristic designs rather than rigorous theoretical analysis. Some works \citep{yoo2024advancing,jiang2024credid,cohen2025watermarking,qu2025provably} divide the generated text into blocks and embeds a watermark bit in each by using different red/green lists to represent the binary values $0$ and $1$. While some studies~\citep{xu2024robust, zamir2024undetectable} predefine a mapping rule or use cryptographic techniques to determine which bit carries the watermark, and the work of \citet{gunn2024undetectable} further extends this idea to image generation model. 
%Most existing multi-bits watermark algorithms are heuristic and rigorous theoretic analysis remains scarce. 
% A few theory-related studies have investigated the tradeoff between different multi-bit watermark performance metrics. 
For theoretical analysis, \citet{boroujeny2024multi} studies the tradeoff between embedding information rate and decoding error probability, by enlarging the message set to increase the embedding rate while designing a distortion-free scheme that preserves the model’s token distribution. However, a unified information-theoretic analysis of multi-bit watermark remains under explored. We propose the first general theoretical framework to study the fundamental limits of distribution information embedding with side information.\looseness=-1

\section{More Discussion on Side Information-based Formulation}
\label{sec:side-information-example}
% \begin{example}[Existing watermarking schemes as special cases]\label{Ex: watermark schemes}
\smallskip
\noindent$\bullet$ \textbf{Green-Red List watermarking}~\citep{kirchenbauer2023watermark}. In this scheme, at each time $t$, the vocabulary $\calV$ is randomly partitioned into a green set $\calG$ and a red set $\calR$, with $|\calG|=\rho|\calV|$ for some $\rho\in(0,1)$. 
This partition can be encoded by a binary side-information vector $\zeta_t\in\{0,1\}^{|\calV|}$ indexed by $x\in\calV$, where $\zeta_t(x)=1$ indicates $x\in\calG$ (and $\zeta_t(x)=0$ indicates $x\in\calR$).

The procedure can be summarized as:
\begin{enumerate}[noitemsep,topsep=0pt,parsep=0pt,partopsep=0pt,leftmargin=*,label=--]
    \item Compute a hash of the previous token using a shared secret: $h(X_{t-1},\key)$, where $h:\calV\times\bbR\to\bbR$.
    \item Use $h(X_{t-1},\key)$ as a seed to draw $\zeta_t$ uniformly from
    $\{\zeta\in\{0,1\}^{|\calV|}:\|\zeta\|_1=\rho|\calV|\}$, which determines $\calG$.
    \item Draw $X_t$ from a modified next-token distribution that upweights green tokens by a logit shift $\delta>0$:
    \setlength{\abovedisplayskip}{5pt}
    \setlength{\belowdisplayskip}{-5pt}
    \setlength{\abovedisplayshortskip}{5pt}
    \setlength{\belowdisplayshortskip}{-5pt}
    \begin{equation}
     \resizebox{.5\hsize}{!}{$
        P_{X_t|x_1^{t-1},\zeta_t}(x)=
        \frac{Q_{X_t|x_1^{t-1}}(x)\exp(\delta\,\mathbbm{1}\{\zeta_t(x)=1\})}
        {\sum_{x\in\calV}Q_{X_t|x_1^{t-1}}(x)\exp(\delta\,\mathbbm{1}\{\zeta_t(x)=1\})}.
     $}
    \end{equation}
\end{enumerate}
% \end{example}

The same side-information perspective also captures several other watermarking schemes in the literature.

\smallskip
\noindent$\bullet$ \textbf{Gumbel-Max watermarking}~\citep{gumbel2023}. 
This method uses the Gumbel-Max trick~\citep{gumbel1954statistical}, where the randomness injected by the Gumbel variables plays the role of side information.
Here, $\zeta_t$ is a $|\calV|$-dimensional vector indexed by $x\in\calV$.
For $t=1,2,\ldots$:
\begin{enumerate}[noitemsep,topsep=0pt,parsep=0pt,partopsep=0pt,leftmargin=*,label=--]
    \item Compute a seed from the previous $n$ tokens and the secret key: $h(X_{t-1}^{t-n},\key)$, with $h:\calV^n\times\bbR\to\bbR$.
    \item Use this seed to sample $\zeta_t$ uniformly from $[0,1]^{|\calV|}$.
    \item Generate the next token via
    \begin{equation}
        X_t=\argmax_{x\in\calV}\ \log Q_{X_t|x_1^{t-1}}(x)-\log\big(-\log \zeta_t(x)\big).
    \end{equation}
\end{enumerate}

\smallskip
\noindent$\bullet$ \textbf{Inverse-transform watermarking}~\citep{kuditipudi2023robust}. 
This scheme treats $\calV$ as $[|\calV|]$ and uses side information consisting of a uniform random variable together with a random permutation.
Specifically, let $\zeta_t=(U_t,\pi_t)$, where $U_t\sim\mathrm{Unif}[0,1]$ and $\pi_t$ is a uniform permutation of $[|\calV|]$, both generated from the shared key.
\begin{enumerate}[noitemsep,topsep=0pt,parsep=0pt,partopsep=0pt,leftmargin=*,label=--]
    \item Using $\key$ as a seed, sample $\{U_t\}_{t=1}^T$ i.i.d.\ from $[0,1]$ and $\{\pi_t\}_{t=1}^T$ i.i.d.\ from the permutation group over $[|\calV|]$.
    \item Sample $X_t$ by inverse transform with respect to the permuted ordering:
    \begin{equation}
        X_t=\pi_t^{-1}\!\left(
        \min\left\{i\in[|\calV|]:
        \sum_{x\in[|\calV|]} Q_{X_t|x_1^{t-1}}(x)\,\mathbbm{1}\{\pi_t(x)\le i\}\ge U_t
        \right\}
        \right),
    \end{equation}
    where $\pi_t^{-1}$ denotes the inverse permutation.
\end{enumerate}

\smallskip
\noindent$\bullet$ \textbf{Adaptive watermarking}~\citep{liu2024adaptive}. 
This approach is structurally similar to Green--Red List watermarking, but replaces the hash with a pretrained neural network.
Concretely, a network $h$ takes a semantic representation $\phi(X_1^{t-1})$ of the generated prefix together with the secret $\key$ and outputs a seed $h(\phi(X_1^{t-1}),\key)$. 
Using this seed, the scheme samples $\zeta_t$ uniformly from
$\{\bv\in\{0,1\}^{|\calV|}:\|\bv\|_1=\rho|\calV|\}$ to determine the green set, and then draws $X_t$ using the same logit-shifted sampling rule as in the Green--Red List method.

\section{Proof of Converse for Theorem \ref{thm: opt ave beta strong}}\label{App: coverse of thm: opt ave beta strong}
First, let us fix a decoder $\gamma\in\calP([0:m]\mid \calX^T\times \calZ^T)$. From the worst-case false-alarm constraint, we have for all $x^T\in \calX^T$,
    \begin{align}
        \alpha\geq \sup_{Q_{X^T}} \beta_0(\gamma,Q_{X^T}\otimes P_{\zeta^T})&\overset{\text{(a)}}{\geq} \sum_{\zeta^T}P_{\zeta^T}(\zeta^T)(1-\gamma(0|x^T,\zeta^T))\\
        &=\sum_{i\in[m]}\sum_{\zeta^T}P_{\zeta^T}(\zeta^T)\gamma(i|x^T,\zeta^T), \label{eq: worst FA constraint}
    \end{align}
    where (a) follows by choosing $Q_{X^T}=\delta_{x^T}$.
    Therefore, there exists a sequence $(\alpha_1(x^T),\ldots, \alpha_m(x^T))$ such that
    \begin{align}
        &\sum_{\zeta^T}P_{\zeta^T}(\zeta^T)\gamma(i|x^T,\zeta^T)\leq\alpha_i(x^T), \quad \forall i, x^T, \quad \text{and}~\sum_{i\in[m]}\alpha_i(x^T)=\alpha. \label{Eq: ith FA constraint}
    \end{align}
        The $j$-th error probability can be rewritten as follows:
    \begin{align}
        \beta_j(\gamma,P_{X^T,\zeta^T|M=j})&=1-\bbE_{\bbP_j}[\gamma(j|X^T,\zeta^T)]\\
        &=1-\sum_{x^T,\zeta^T}P_{\zeta^T}(\zeta^T)P_{X^T|\zeta^T,M}(x^T|\zeta^T,j)\gamma(j|x^T,\zeta^T),\label{eq: j-th error decomp ass 1}
    \end{align}
    and thus the average detection error probability is given by
    \begin{equation}
        \bar{\beta}(\gamma, (\bbP_j)_{j\in[m]})=1-\frac{1}{m}\sum_{j\in[m]}\sum_{x^T,\zeta^T}P_{\zeta^T}(\zeta^T)P_{X^T|\zeta^T,M}(x^T|\zeta^T,j)\gamma(j|x^T,\zeta^T).\label{eq: ave wm fail prob ass 1}
    \end{equation}

To derive a lower bound of $\bar{\beta}(\gamma, (\bbP_j)_{j\in[m]})$, we first bound the second term of \eqref{eq: ave wm fail prob ass 1} as follows: for all $j\in[m]$ and $x^T\in\calX^T$,
\begin{align}
    &\sum_{\zeta^T}P_{\zeta^T}(\zeta^T)\underbrace{P_{X^T|\zeta^T,M}(x^T|\zeta^T,j)}_{\leq 1}\gamma(j|x^T,\zeta^T)\\
    &\leq \sum_{\zeta^T}P_{\zeta^T}(\zeta^T)\gamma(j|x^T,\zeta^T) \overset{\text{(a)}}{\leq} \alpha_j(x^T),\label{eq: j-th ub 1 ass 1}
\end{align}
where (a) follows from \eqref{Eq: ith FA constraint},
and then
\begin{align}
    \frac{1}{m}\sum_{j\in[m]}\sum_{\zeta^T}P_{\zeta^T}(\zeta^T)\gamma(j|x^T,\zeta^T)\leq \frac{1}{m}\sum_{j\in[m]}\alpha_j(x^T)=\frac{\alpha}{m}.\label{eq: ave ub 1 ass 1}
\end{align}
On the other hand, since $\mathbbm{1}\{\gamma(x^T,\zeta^T)=j\}\leq 1$, we also have: for all $j\in[m]$ and $x^T\in\calX^T$,
\begin{align}
    &\sum_{\zeta^T}P_{\zeta^T}(\zeta^T)P_{X^T|\zeta^T,M}(x^T|\zeta^T,j)\gamma(j|x^T,\zeta^T)\nn\\
    &\leq \sum_{\zeta^T}P_{\zeta^T}(\zeta^T)P_{X^T|\zeta^T,M}(x^T|\zeta^T,j)=P_{X^T|M}(x^T|j)\overset{\text{(a)}}{=}P_{X^T}(x^T), \label{eq: j-th ub 2 ass 1}
\end{align}
where (a) follows from Assumption \ref{Ass: strong secrecy}.
Hence, for all $x^T\in\calX^T$, the following inequality also holds:
\begin{align}
    \frac{1}{m}\sum_{j\in[m]}\sum_{\zeta^T}P_{\zeta^T}(\zeta^T)P_{X^T|\zeta^T,M}(x^T|\zeta^T,j)\gamma(j|x^T,\zeta^T)\leq P_{X^T}(x^T). \label{eq: ave ub 2 ass 1}
\end{align}
Combining \eqref{eq: ave ub 1 ass 1} and \eqref{eq: ave ub 2 ass 1}, we lower bound $\bar{\beta}$ as follows:
\begin{align}
    \bar{\beta}(\gamma, (\bbP_j)_{j\in[m]})\geq 1-\sum_{x^T\in\calX^T} \left(P_{X^T}(x^T)\wedge \frac{\alpha}{m}\right)=\sum_{x^T\in\calX^T}\left(P_{X^T}(x^T)-\frac{\alpha}{m}\right)_+.
\end{align}
As this lower bound holds for all $P_{X^T}$ satisfying the distortion constraint $\rvD(P_{X^T},Q_{X^T})\leq d$, it is minimized when $P_{X^T}$ is chosen as follows
\begin{equation}
    P_{X^T}^*=\argmin_{P_{X^T}:\rvD(P_{X^T},Q_{X^T})\leq d}\sum_{x^T\in\calX^T}\left(P_{X^T}(x^T)-\frac{\alpha}{m}\right)_+,
\end{equation}
and eventually
\begin{equation}
    \bar{\beta}(\gamma, (\bbP_j)_{j\in[m]})\geq \min_{P_{X^T}:\rvD(P_{X^T},Q_{X^T})\leq d}\sum_{x^T\in\calX^T}\left(P_{X^T}(x^T)-\frac{\alpha}{m}\right)_+. \label{Eq: lower bd for ave beta_j ass 1}
\end{equation}
We observe that the lower bound \eqref{Eq: lower bd for ave beta_j ass 1} is independent of $\gamma$. Thus, the lower bound also holds for the optimal value of the optimization problem \eqref{Eq: opt-O}.

For the optimization problem \eqref{Eq: opt-i}, we apply a similar method to derive a lower bound for $\beta_j$.
By combining \eqref{eq: j-th ub 1 ass 1} and \eqref{eq: j-th ub 2 ass 1}, for all $j\in[m]$, we have
\begin{align}
    \beta_j(\gamma,P_{X^T,\zeta^T|M=j})&\geq  1- \sum_{x^T} \bigg(P_{X^T}(x^T)\wedge \alpha_j(x^T)\bigg)\\
    &=\sum_{x^T} \bigg(P_{X^T}(x^T)-\alpha_j(x^T)\bigg)_+\\
    &\geq \min_{P_{X^T}: \rvD(P_{X^T},Q_{X^T})\leq d}\sum_{x^T} \bigg(P_{X^T}(x^T)-\alpha_j(x^T)\bigg)_+.
\end{align}
Among all possible sequences $(\alpha_1(x^T),\ldots, \alpha_m(x^T))$ that sum up to $\alpha$, the one that minimizes the lower bound for $\max_{j\in[m]}\beta_j(\gamma,P_{X^T,\zeta^T|M=j})$ is $(\frac{\alpha}{m},\ldots, \frac{\alpha}{m})$. The proof is as follows:
\begin{align}
    \max_{j\in[m]}\beta_j(\gamma,P_{X^T,\zeta^T|M=j})&\geq \max_{j\in[m]}\min_{P_{X^T}: \rvD(P_{X^T},Q_{X^T})\leq d}\sum_{x^T} \bigg(P_{X^T}(x^T)-\alpha_j(x^T)\bigg)_+ \\
    &\overset{(a)}{\geq} \min_{P_{X^T}: \rvD(P_{X^T},Q_{X^T})\leq d} \sum_{x^T} \bigg(P_{X^T}(x^T)-\frac{\alpha}{m}\bigg)_+ , \label{Eq: lower bd for max beta_j strong}
\end{align}
where (a) holds with equality when $\alpha_j(x^T)=\frac{\alpha}{m}$ for all $j\in[m]$. 
Similarly, the lower bound \eqref{Eq: lower bd for max beta_j strong} is independent of $\gamma$. Thus, the lower bound also holds for the optimal value of the optimization problem \eqref{Eq: opt-i}.

In particular, if we choose the distortion function to be the total variation distance $\rvD_\TV$, the lower bound can be further simplified as follows.

For any $P_{X^T}$ satisfying the distortion constraint, define
\begin{equation}
  F(P_{X^T})\coloneqq\sum_{x^T} \bigg(P_{X^T}(x^T)-\frac{\alpha}{m}\bigg)_+.
\end{equation}
Define
\begin{equation}
    E(Q_{X^T};\frac{\alpha}{m})\coloneqq \sum_{x^T}(Q_{X^T}(x^T)-\frac{\alpha}{m})_+,\quad
S(Q_{X^T};\frac{\alpha}{m})\coloneqq \sum_{x^T}(\frac{\alpha}{m}-Q_{X^T}(x^T))_+.
\end{equation}
Let $\calA\coloneqq\{x^T: Q_{X^T}(x^T)>\frac{\alpha}{m}\}$. We have
\begin{align}
    F(P_{X^T})\geq \sum_{x^T:x^T\in\calA} \bigg(P_{X^T}(x^T)-\frac{\alpha}{m}\bigg)_+ &\geq \sum_{x^T:x^T\in\calA} \bigg(P_{X^T}(x^T)-\frac{\alpha}{m}\bigg)\\
    &=\sum_{x^T:x^T\in\calA} \bigg(P_{X^T}(x^T)-Q_{X^T}(x^T)+Q_{X^T}(x^T)-\frac{\alpha}{m}\bigg)\\
    &=E(Q_{X^T};\frac{\alpha}{m})+\sum_{x^T:x^T\in\calA}(P_{X^T}(x^T)-Q_{X^T}(x^T)),
\end{align}
which leads to
\begin{align}
    E(Q_{X^T};\frac{\alpha}{m})-F(P_{X^T})&\leq \sum_{x^T:x^T\in\calA}(Q_{X^T}(x^T)-P_{X^T}(x^T))\\
    &\leq \sum_{x^T:x^T\in\calA}(Q_{X^T}(x^T)-P_{X^T}(x^T))_+\\
    &\leq \sum_{x^T}(Q_{X^T}(x^T)-P_{X^T}(x^T))_+\leq d.
\end{align}
Thus, we have one lower bound for $F(P_{X^T})$, i.e.,
\begin{equation}
    F(P_{X^T})\geq  E(Q_{X^T};\frac{\alpha}{m})-d. \label{eq: F(P) lb 1}
\end{equation}

On the other hand, we can derive another lower bound for $F(P_{X^T})$ using the fact that $F(P_{X^T})=1-\sum_{x^T}(P_{X^T}(x^T)\wedge \frac{\alpha}{m})$. For each $x^T\in\calX^T$,
\begin{itemize}
    \item If $Q_{X^T}(x^T)\geq\frac{\alpha}{m}$, then $Q_{X^T}(x^T)\wedge \frac{\alpha}{m}=\frac{\alpha}{m}$ and $P_{X^T}(x^T)\wedge \frac{\alpha}{m}\leq \frac{\alpha}{m}=Q_{X^T}(x^T)\wedge \frac{\alpha}{m}$;
    \item If $Q_{X^T}(x^T)<\frac{\alpha}{m}$, then $Q_{X^T}(x^T)\wedge \frac{\alpha}{m}=Q_{X^T}(x^T)$ and $P_{X^T}(x^T)\wedge \frac{\alpha}{m}\leq \frac{\alpha}{m}=Q_{X^T}(x^T)+(\frac{\alpha}{m}-Q_{X^T}(x^T))=Q_{X^T}(x^T)\wedge \frac{\alpha}{m}+(\frac{\alpha}{m}-Q_{X^T}(x^T))_+$.
\end{itemize}
As a result, we have
\begin{align}
    P_{X^T}(x^T)\wedge \frac{\alpha}{m}&\leq Q_{X^T}(x^T)\wedge \frac{\alpha}{m}+(\frac{\alpha}{m}-Q_{X^T}(x^T))_+,\\
    \text{and}~~1-F(P_{X^T})=\sum_{x^T}(P_{X^T}(x^T)\wedge \frac{\alpha}{m})&\leq \sum_{x^T}(Q_{X^T}(x^T)\wedge \frac{\alpha}{m})+S(Q_{X^T};\frac{\alpha}{m}).
\end{align}
Therefore, $F(P_{X^T})$ has another lower bound as follows:
\begin{align}
    F(P_{X^T})&\geq 1-\sum_{x^T}(Q_{X^T}(x^T)\wedge \frac{\alpha}{m})-S(Q_{X^T};\frac{\alpha}{m})\\
    &=E(Q_{X^T};\frac{\alpha}{m})-S(Q_{X^T};\frac{\alpha}{m}).  \label{eq: F(P) lb 2}
\end{align}
From \eqref{eq: F(P) lb 1} and \eqref{eq: F(P) lb 1} and the fact that $F(P_{X^T})\geq 0$, we finally have
\begin{equation}
    F(P_{X^T})\geq \bigg(E(Q_{X^T};\frac{\alpha}{m})-\bigg(d \wedge S(Q_{X^T};\frac{\alpha}{m})\bigg)\bigg)_+. \label{eq: F(P) final lb}
\end{equation}
There exists a distribution $P^*_{X^T}$ that can achieve this lower bound in \eqref{eq: F(P) final lb}. Intuitively, constructing $P^*_{X^T}$ is equivalent to the process of moving the mass of $Q_{X^T}$ above $\alpha$ to below $\alpha$, under some constraints. 

Define
\begin{align}
    \delta &\coloneqq \min\bigg\{d,\, S(Q_{X^T};\frac{\alpha}{m}),\, E(Q_{X^T};\frac{\alpha}{m})\bigg\}, \\
    \calA&\coloneqq\{x^T: Q_{X^T}(x^T)>\frac{\alpha}{m}\}, \quad \calB\coloneqq\{x^T: Q_{X^T}(x^T)\leq\frac{\alpha}{m}\}.
\end{align}
Choose nonnegative transfers $\tau(x^T,y^T)\ge 0$ satisfying
$$\sum_{y^T\in \calB}\tau(x^T,y^T)\le Q_{X^T}(x^T)-\frac{\alpha}{m}, \quad
\sum_{x^T\in \calA}\tau(x^T,y^T)\le \frac{\alpha}{m}-Q_{X^T}(x^T),\quad
\sum_{x^T\in \calA}\sum_{y^T\in \calB}\tau(x^T,y^T)=\delta.$$ 
Such $\tau$ can always be constructed using greedy algorithms. Then the optimal $P^*_{X^T}$ that achieves the lower bound in \eqref{eq: F(P) final lb} is given by
\begin{align}
    P^*_{X^T}(x^T)=\begin{cases}
        Q_{X^T}(x^T)-\sum_{y^T\in \calB}\tau(x^T,y^T), & x^T\in\calA,\\
        Q_{X^T}(x^T)+\sum_{y^T\in \calA}\tau(y^T,x^T), & x^T\in\calB.
    \end{cases} 
\end{align}
Note that $P^*_{X^T}$ is a valid probability pass function and satisfies $\rvD_\TV(P^*_{X^T},Q_{X^T})=\delta\leq d$.
By such construction, $P^*_{X^T}(x^T)\leq \frac{\alpha}{m}$ for all $x^T\in \calB$ and  $P^*_{X^T}(x^T)> \frac{\alpha}{m}$ for all $x^T\in \calA$. Then the induced $F(P^*_{X^T})$ is
\begin{align}
    F(P^*_{X^T})&=\sum_{x^T}(P^*_{X^T}(x^T)-\frac{\alpha}{m})_+ \\
    &= \sum_{x^T:x^T\in\calA}(P^*_{X^T}(x^T)-\frac{\alpha}{m})=\sum_{x^T:x^T\in\calA}( Q_{X^T}(x^T)-\sum_{y^T\in \calB}\tau(x^T,y^T)-\frac{\alpha}{m})\\
    &=E(Q_{X^T};\frac{\alpha}{m})-\delta.
\end{align}
If $\delta=E(Q_{X^T};\frac{\alpha}{m})$, then $F(P^*_{X^T})=0$; otherwise, $F(P^*_{X^T})=E(Q_{X^T};\frac{\alpha}{m})-\bigg(d \wedge S(Q_{X^T};\frac{\alpha}{m})\bigg)$. Thus, $P^*_{X^T}$ achieves the lower bound in \eqref{eq: F(P) final lb}, proving optimality. In conclusion, under the distortion constraint defined with total variation distance, the lower bound for the detection error probability is given by the lower bound in \eqref{eq: F(P) final lb}.

\section{Proof of Achievability for Theorem \ref{thm: opt ave beta strong}}\label{App: pf of achievability thm1}

\subsection{Randomized Decoding (Proof of Theorem~\ref{thm: ach rand scheme})}
We define a stochastic grouping kernel $\phi:\calX^T\to[K]$ (with $K\ge m$) satisfying $\sum_{k\in[K]}\phi(k|x^T)=1$ for all $x^T$, which induces group probabilities
\[
P_G(k)\;\coloneqq\;\Pr(G=k)
=\sum_{x^T} P^*_{X^T}\phi(k|x^T).
\]
We aim to construct a kernel $\phi$ such that there are at least $m$ groups, denoted by $k_1,k_2,\ldots,k_m$, with probability $P_G(k_l)\geq \frac{\alpha}{m}$ for all $l\in[m]$, while the overflow mass above $\frac{\alpha}{m}$ matches the lower bound in \eqref{eq: dec error opt}:
\[
\sum_{k\in[K]}(P_G(k)-\frac{\alpha}{m})_+=\sum_{x^T} (P^*_{X^T}(x^T)-\frac{\alpha}{m})_+ \triangleq \LB(Q_{X^T},\alpha,d,m).
\]
Note that the lower bound $\LB(Q_{X^T},\alpha,d,m)$ (cf.~\eqref{eq: dec error opt}) is non-decreasing in $m$. Therefore, to ensure such a kernel $\phi$ exists, the message set size $m$ cannot exceed:
\begin{equation}
  m^*\coloneqq \max\{m\in[|\calX|^T]:\LB(Q_{X^T},\alpha,d,m)\leq 1-\alpha\}.
\end{equation}
For any $m\le m^*$, there exists a stochastic grouping kernel $\phi$ achieving the desired properties. Under this condition, the total probability mass of $P^*_{X^T}$ below $\frac{\alpha}{m}$ exceeds $\alpha$, which ensures that the mass can be redistributed via $\phi$ to form $m$ groups each with probability at least $\alpha/m$.  

As a simple illustration, if $P^*_{X^T}$ is a uniform distribution (i.e., the maximum-entropy distribution on $|\calX|^T$), then $m^*=2^{\rvH(P^*_{X^T})}=|\calX|^T$. In the following, we present an example to illustrate the existence of such a kernel $\phi$ for any $P^*_{X^T}$.

\paragraph{Example construction of $P_G$ and $\phi$}
Define a target group PMF $(p_k)_{k=1}^K$ by
\begin{equation}
    p_1=\cdots=p_{m-1}=\frac{\alpha}{m}, \quad p_m=c+\LB, \quad \sum_{k=m+1}^K p_k=1-\alpha-\LB,
\end{equation}
with each $P_G(k)\in[0,\frac{\alpha}{m})$ for $k\ge m+1$.
Once we have the target PMF $(p_k)_{k=1}^K$, there are many $\phi$’s that induce it. The simplest is the ``independent'' kernel:
\[
\phi(k\mid x^T) \;\triangleq\; p_k,\qquad \forall x^T,
\]
which means moving $p_k$ fraction of each $P^*_{X^T}$ to the group $k$.
Then
\[
P_G(k)=\sum_{x^T}P^*_{X^T}(x^T)\phi(k\mid x^T)= p_k\sum_{x^T}P^*_{X^T}(x^T)= p_k, \quad \forall k\in[K].
\]

\paragraph{Existence of optimal decoder} For simplicity, we let the auxiliary random variable $\zeta^T$ take values in a finite alphabet $\calZ^T=[K]\cup\{\widetilde{\zeta^T}\}$, whose cardinality may depend on $T$, but which does not represent a length-$T$ Cartesian
product. Define $\calG_{\alpha/m}\coloneqq\{k\in[K]: P_G(k)\geq \frac{\alpha}{m}\}$. Let $h: \calG_{\alpha/m}\times \calG_{\alpha/m} \to [|\calG_{\alpha/m}|]$ be a function such that for every $k\in \calG_{\alpha/m}$, $h(k,\cdot)$ is a bijection, and for every $\zeta^T\in \calG_{\alpha/m}$, $h(\cdot,\zeta^T)$ is a bijection. A Latin square matrix satisfies such structure. %Given any $j\in[|\calG_{\alpha/m}|]$, we denote the inverse of $h$ by $h_j^{-1}$.

Based on the stochastic group kernel $\phi$, we choose the randomized decoder as
\begin{equation}
    \gamma^*(j|x^T,\zeta^T)=\sum_{k\in[K]}\phi(k|x^T)\gamma_G(j|k,\zeta^T)%\begin{cases}
        % \frac{\phi(h_j^{-1}(\zeta^T)|x^T)}{\sum_{j=1}^m \phi(h_j^{-1}(\zeta^T)|x^T)}, & \text{if}~\zeta^T\in \calG_{\alpha/m} ~~\text{and}~~j\leq m\\
        % 1-\mathbbm{1}\{g(\zeta^T)=\phi^*(x^T)\}, & \text{if}~\zeta^T\in \calZ_<^T,\\
        % 0, & \text{otherwise}.%\text{if}~\zeta^T=\widetilde{\zeta^T},\\
    % \end{cases}
    \label{eq: opt sto dec}
\end{equation}
where $\gamma_G:[K]\times \calZ^T \to [0:m]$ is a group-level decoder
\begin{equation}
    \gamma_G(j|k,\zeta^T)=\mathbbm{1}\{h(k,\zeta^T)=j\}, \quad \gamma_G(0|k,\widetilde{\zeta^T})=1.
\end{equation}
Given $(x^T,\zeta^T)$, the decoder samples $\widehat G\sim \phi(\cdot\mid x^T)$, and then outputs $\hat M\sim \gamma_G(\cdot\mid \widehat G,\zeta^T)$.

\paragraph{Existence of optimal watermark embedding:}
The corresponding conditional  distribution $P^*_{\zeta^T|X^T,M}$ is constructed as follows: for all $j\in[m]$,
\begin{align}
P^*_{\zeta^T|X^T,M}(\zeta^T|x^T,j)&=\sum_{k\in[K]}P^*_{\zeta^T,G|X^T,M}(\zeta^T,k|x^T,j)=\sum_{k\in[K]}P^*_{\zeta^T|G,M}(\zeta^T|k,j)\phi(k|x^T),
\end{align}
where for all $k\in[K]$
\begin{align}
    P^*_{\zeta^T|G,M}(\zeta^T|k,j)=\begin{cases}
        1\wedge \frac{\alpha}{m P_G(k)}, ~ &\text{if}~ h(k,\zeta^T)=j,\\
       1, &\text{if}~\zeta^T\in [K]\setminus\calG_{\alpha/m}~~\text{and}~~ \zeta^T=k \\
        (1-\frac{\alpha}{m P_G(k)})_+, &\text{if}~ \zeta^T=\widetilde{\zeta^T},\\
        0, & \text{otherwise}.
    \end{cases}\label{eq: opt cond distr}
\end{align}
We can verify that $P^*_{\zeta^T|M=j}$ is the same for all $j\in[m]$ as follows:
\begin{align}
    P^*_{\zeta^T|M}(\zeta^T|j)&=\sum_{k\in[K]}P^*_{\zeta^T|G,M}(\zeta^T|k,j)P_{G}(k)\\
    &=\begin{cases}
        \frac{\alpha}{m}, &\text{if}~ \zeta^T\in \calG_{\alpha/m},\\
        P_G(\zeta^T), &\text{if}~ \zeta^T\notin \calG_{\alpha/m}\cup\{\widetilde{\zeta^T}\},\\
        \sum_{k\in[K]}(P_G(k)-\frac{\alpha}{m})_+, &\text{if}~ \zeta^T=\widetilde{\zeta^T},
    \end{cases}\\
    &\triangleq P^*_{\zeta^T}(\zeta^T).
\end{align}

\paragraph{Achievable Errors:}
For a randomized decoder $\gamma(\cdot|x^T,\zeta)$, the $j$-th error  probability is
\[
\beta_j(\gamma,\bbP_j)
=1-\sum_{x^T,\zeta^T}P_{X^T,\zeta|M=j}(x^T,\zeta^T)\,\gamma(j|x^T,\zeta^T).
\]

Given $(x^T,\zeta^T)$, the decoder samples $\widehat G\sim \phi(\cdot\mid x^T)$, and then outputs $\hat M\sim \gamma_G(\cdot\mid \widehat G,\zeta^T)$. Then the success probability is
\[\Pr(\hat M=j\mid M=j)
= \sum_{k} P_G(k)\sum_{\zeta^T}P_{\zeta^T\mid G,M}(\zeta^T\mid k,j)\,\gamma_G(j\mid k,\zeta^T)=\sum_{k\in[K]}(P_G(k)\wedge\frac{\alpha}{m}).
\]

The induced $j$-th error probability is given by
\begin{align}
    \beta_j(\gamma,\bbP_j)&=1-\Pr(\hat M=j\mid M=j)\\
    &=1-\sum_{k\in[K]}(P_G(k)\wedge\frac{\alpha}{m})\\
    &=\sum_{k\in[K]}(P_G(k)-\frac{\alpha}{m})_+\\
    &=\underbrace{\sum_{x^T }(P^*_{X^T}(x^T)-\frac{\alpha}{m})_+}_{\text{lower bound in Theorem \ref{thm: opt ave beta strong}}}.
\end{align}

The induced false-alarm error is derived as follows. For all $x^T\in\calX^T$, we have
\begin{align}
    \sum_{j\in[m]}\sum_{\zeta^T}P^*_{\zeta^T}(\zeta^T)\gamma^*(j|x^T,\zeta^T)= \sum_{j\in[m]}\sum_{\zeta^T}P^*_{\zeta^T}(\zeta^T)\sum_{k\in[K]}\phi(k|x^T)\gamma_G(j|k,\zeta^T)=m\cdot \frac{\alpha}{m}=\alpha.
\end{align}
Thus, as any $Q_{X^T}$ can be represented by a linear combination of $\{\delta_{x^T}\}_{x^T\in\calX^T}$, the worst-case false-alarm error is given by
\begin{align}
    \beta_0(\gamma,Q_{X^T}\otimes P_{\zeta^T})=\sup_{Q_{X^T}}\sum_{j\in[m]}\sum_{x^T,\zeta^T}Q_{X^T}(x^T)P^*_{\zeta^T}(\zeta^T)\gamma^*(j|x^T,\zeta^T)=\alpha.
\end{align}
The optimality of the construction is thus proved.

\subsection{Deterministic Decoding (Proof of Theorem~\ref{thm: ach det scheme main})}\label{App: determ dec proof}
Given any source distribution $Q_{X^T}$, any message set size $m\le |\calX|^T$, and any false-alarm constraint $\alpha \in(0,1)$, the optimal joint distribution of the watermarked sequence $X^T$ is given by
\begin{equation}
    P^*_{X^T}\coloneqq \argmin_{P_{X^T}: \rvD(P_{X^T},Q_{X^T})\leq d} \sum_{x^T} \bigg(P_{X^T}(x^T)-\frac{\alpha}{m}\bigg)_+.
\end{equation}
Then the next step is to find the conditional  distribution $P^*_{\zeta^T|X^T,M}$ and the detector $\gamma^*$ such that detection error lower bound is achieved and the worst-case false-alarm constraint is satisfied. 

We define a deterministic grouping function $\phi:\calX^T\to[K]$ (with $K\ge m$) that assigns
each symbol $x^T\in\calX^T$ to a group index $k\in[K]$, thereby inducing group probabilities
\[
P_G(k)\;\coloneqq\;\Pr(G=k)
=\sum_{x^T:\,\phi(x^T)=k} P^*_{X^T}(x^T).
\]
We impose the additional constraint that any symbol
$x^T$ with $P^*_{X^T}(x^T)\ge \frac{\alpha}{m}$ forms a singleton group, i.e.,
\[
P^*_{X^T}(x^T)\ge \frac{\alpha}{m} \;\;\Rightarrow\;\; \phi(x^T)\neq \phi(\tilde x^T)\ \text{for all }\tilde x^T\neq x^T.
\]
Let $\calS_{\alpha/m}\coloneqq\{x^T: P^*_{X^T}(x^T)\ge \frac{\alpha}{m}\}$ and $\calG_{\alpha/m}\coloneqq\{k\in[K]: P_G(k)\geq \frac{\alpha}{m}\}$.
We define $\phi^\star$ as a solution to the optimization problem
\begin{align}
&\min_{\phi:\calX^T\to[K]} \quad
 \sum_{k=1}^K \bigl(P_G(k)-\frac{\alpha}{m}\bigr)_+ \tag{P3} \label{eq: opt phi}\\
\text{s.t.}\quad &
\bigl|\calG_{\alpha/m}\bigr|\ge m,\quad \text{every}~~x^T\in\calS_{\alpha/m} \text{ is assigned to a singleton group}.
\end{align}
The feasible set is non-empty if $m\in\calM\coloneqq \{m\in [|\calX|^T]: \sum_{x^T\notin \calS_{\alpha/m}}P^*_{X^T}(x^T)\geq (1-\frac{|\calS_{\alpha/m}|}{m})\alpha\}$. This means it is possible to redistribute the mass below $\frac{\alpha}{m}$ to at least $m-|\calS_{\alpha/m}|$ groups with probability exceeding $\frac{\alpha}{m}$.
% Since $\sum_{k=1}^K P_G(k)=1\ge m\cdot \frac{\alpha}{m}=\alpha$, the feasible set is nonempty.
Moreover, $\phi^\star$ exists because the set of deterministic groupings of a finite alphabet
is finite. Intuitively, this optimization seeks to partition the probability mass of
$P^*_{X^T}$ into at least $m$ groups of mass at least $\frac{\alpha}{m}$ while minimizing the total overflow
above $\frac{\alpha}{m}$ across all groups. A feasible grouping can be constructed via a greedy
procedure, while $\phi^\star$ denotes an overflow-minimizing solution.

% More specifically, consider a greedy ``fill-to-$\frac{\alpha}{m}$'' procedure: 1) Sort points in $\calX^T$ by probability $P^*_{X^T}$ in a descending order; 2) Iterate through symbols $x^T$ and keep adding their masses to the current group For each $x^T$, if $P^*_{X^T}(x^T)\geq \frac{\alpha}{m}$, let $\phi(x^T)=order(x^T)$ and increase $k$ by $1$; 

\paragraph{Optimal Decoder:} Let the auxiliary alphabet be the union of three disjoint sets: $\calZ^T=\calZ_{\ge}^T\cup \calZ_<^T \cup \{\widetilde{\zeta^T}\}$, where $|\calZ_{\ge}^T|=|\calG_{\alpha/m}|$ and $|\calZ_<^T|=K-|\calG_{\alpha/m}|$. Based on the optimized function $\phi^*$, we choose the decoder as
\begin{equation}
    \gamma^*(x^T,\zeta^T)=\begin{cases}
        j, & \text{if}~\zeta^T\in \calZ_\ge^T ~~\text{and}~~h(\phi^*(x^T),\zeta^T)=j\leq m\\
        % 1-\mathbbm{1}\{g(\zeta^T)=\phi^*(x^T)\}, & \text{if}~\zeta^T\in \calZ_<^T,\\
        0, & \text{otherwise}.%\text{if}~\zeta^T=\widetilde{\zeta^T},\\
    \end{cases}\label{eq: opt dec}
\end{equation}
where $h: \calG_{\alpha/m}\times \calZ_\ge^T \to [|\calG_{\alpha/m}|]$ is a function such that for every $k\in \calG_{\alpha/m}$, $h(k,\cdot)$ is a bijection, and for every $\zeta^T\in \calZ_\ge^T$, $h(\cdot,\zeta^T)$ is a bijection. A Latin square matrix satisfies such structure.

\paragraph{Optimal Watermark Embedding:} The corresponding conditional  distribution $P^*_{\zeta^T|X^T,M}$ is constructed as follows: for all $j\in[m]$,
\begin{align}
P^*_{\zeta^T|X^T,M}(\zeta^T|x^T,j)&=\sum_{k\in[K]}P^*_{\zeta^T,G|X^T,M}(\zeta^T,k|x^T,j)=\sum_{k\in[K]}P^*_{\zeta^T|G,M}(\zeta^T|k,j)P_{G|X^T}(k|x^T)\\
&=P^*_{\zeta^T|G,M}(\zeta^T|\phi^*(x^T),j),
\end{align}
where for all $k\in[K]$
\begin{align}
    P^*_{\zeta^T|G,M}(\zeta^T|k,j)=\begin{cases}
        1\wedge \frac{\alpha}{m P_G(k)}, ~ &\text{if}~ h(k,\zeta^T)=j,\\
       1, &\text{if}~\zeta^T\in\calZ_<^T ~~\text{and}~~ g(\zeta^T)=k\\
        (1-\frac{\alpha}{m P_G(k)})_+, &\text{if}~ \zeta^T=\widetilde{\zeta^T},\\
        0, & \text{otherwise},
    \end{cases}
\end{align}
and $g:\calZ_<^T\to [K]\setminus\calG_{\alpha/m}$ is a bijective function.
We can verify that $P^*_{\zeta^T|M=j}=P^*_{\zeta^T}$ for all $j\in[m]$ as follows:
\begin{align}
    P^*_{\zeta^T|M}(\zeta^T|j)&=\sum_{k\in[K]}P^*_{\zeta^T|G,M}(\zeta^T|k,j)P_{G}(k)\\
    &=\begin{cases}
        \frac{\alpha}{m}, &\text{if}~ \zeta^T\in \calZ_\ge^T,\\
        P_G(g(\zeta^T)), &\text{if}~ \zeta^T\in \calZ_<^T,\\
        \sum_{k\in[K]}(P_G(k)-\frac{\alpha}{m})_+, &\text{if}~ \zeta^T=\widetilde{\zeta^T},
    \end{cases}\\
    &\triangleq P^*_{\zeta^T}(\zeta^T),
\end{align}
which is the same for all $j\in[m]$.

\paragraph{Achievable Errors:}
The induced $j$-th error probability is given by
\begin{align}
    \beta_j(\gamma,\bbP_j)&=1-\sum_{x^T,\zeta^T}P^*_{X^T}(x^T)P^*_{\zeta^T|X^T,M}(\zeta^T|x^T,j)\mathbbm{1}\{\gamma^*(x^T,\zeta^T)=j\}\\
    &=1-\sum_{x^T,\zeta^T}P^*_{X^T}(x^T)P^*_{\zeta^T|G,M}(\zeta^T|\phi^*(x^T),j)\mathbbm{1}\{\gamma^*(x^T,\zeta^T)=j\}\\
    &=1-\sum_{k\in[K]}\sum_{\zeta^T}\sum_{x^T:\phi^*(x^T)}P^*_{X^T}(x^T)P^*_{\zeta^T|G,M}(\zeta^T|k,j)\mathbbm{1}\{\gamma^*(x^T,\zeta^T)=j\}\\
    &=1-\sum_{k\in[K]}\sum_{\zeta^T}P_G(k)P^*_{\zeta^T|G,M}(\zeta^T|k,j)\mathbbm{1}\{h(k,\zeta^T)=j\}\\
    &=1-\sum_{k\in[K]}(P_G(k)\wedge\frac{\alpha}{m})\\
    &=\sum_{k\in[K]}(P_G(k)-\frac{\alpha}{m})_+\\
    &=\underbrace{\sum_{x^T\in\calS_{\alpha/m} }(P^*_{X^T}(x^T)-\frac{\alpha}{m})_+}_{\text{lower bound in Theorem \ref{thm: opt ave beta strong}}} + \underbrace{\sum_{k\in\calG_{\alpha/m}\setminus \phi^*(\calS_{\alpha/m})}(P_G(k)-\frac{\alpha}{m})_+}_{\coloneqq \varepsilon_{\text{res}}~~ \text{(residual error)}}, \label{Eq: res error}
\end{align}
where $\phi^*(\calS_{\alpha/m})$ denote the collection of the singleton groups. 
Since $\phi^*$ is optimized as an optimizer of \eqref{eq: opt phi}, the residual error is minimized and may vanish for certain generative distributions $Q_{X^T}$. For example, when the distortion level $d=0$ so that $P^*_{X^T}=Q_{X^T}$, and when all probability masses satisfy $Q_{X^T}(x^T)\geq \frac{\alpha}{m}$, then the residual error $\varepsilon_{\text{res}}=0$. In this case, $j$-th error probability $\beta_j(\gamma,\bbP_j)$ reaches the lower bound in Theorem \ref{thm: opt ave beta strong}. 

The induced false-alarm error is derived as follows. For all $x^T\in\calX^T$, we have
\begin{align}
    \sum_{\zeta^T}P^*_{\zeta^T}(\zeta^T)\mathbbm{1}\{\gamma^*(x^T,\zeta^T)\ne 0\}= \sum_{j\in[m]}\sum_{\zeta^T}P^*_{\zeta^T}(\zeta^T)\mathbbm{1}\{\gamma^*(x^T,\zeta^T)=j\}=m\cdot \frac{\alpha}{m}=\alpha.
\end{align}
Thus, as any $Q_{X^T}$ can be represented by a linear combination of $\{\delta_{x^T}\}_{x^T\in\calX^T}$, the worst-case false-alarm error is given by
\begin{align}
    \beta_0(\gamma,Q_{X^T}\otimes P_{\zeta^T})=\sup_{Q_{X^T}}Q_{X^T}\otimes P_{\zeta^T}(\gamma^*(x^T,\zeta^T)\ne 0)=\alpha.
\end{align}

% \section{Proof of Lemma \ref{Lem: distortion asymp}}\label{App: pf of Lem: distortion asymp}
% By the chain rule for KL,
% \[
% \rvD_\KL(P_{X^T}\|Q_{X^T})
% =\sum_{t=1}^T \bbE_{P}\!\left[\rvD_\KL\!\big(P_{X_t|X^{t-1}}(\cdot|X^{t-1})\ \|\ Q_{X_t|X^{t-1}}(\cdot|X^{t-1})\big)
% \right].
% \]
% Divide by $T$ and take $\limsup:
% \overline \rvD_\KL(P\|Q)
% \triangleq \limsup_{T\to\infty}\frac{1}{T}\rvD_\KL(P_{X^T}\|Q_{X^T})
% \le \lim_{T\to\infty}\frac{d}{T}=0$.
% Hence $\overline \rvD_\KL(P\|Q)=0$. The integrand in the chain rule is nonnegative, so $\overline \rvD_\KL(P\|Q)=0$ implies
% $$D\!\big(P_{X_0|X_{-\infty}^{-1}}(\cdot|X_{-\infty}^{-1})\ \|\ Q_{X_0|X_{-\infty}^{-1}}(\cdot|X_{-\infty}^{-1})\big)=0
% \quad P\text{-a.s.},$$
% and therefore
% $$P_{X_0|X_{-\infty}^{-1}}(\cdot|X_{-\infty}^{-1})
% =
% Q_{X_0|X_{-\infty}^{-1}}(\cdot|X_{-\infty}^{-1})
% \quad P\text{-a.s.}$$
% Since both $P$ and $Q$ are stationary, equality of the one-step predictive kernels (given the entire past) implies equality of the induced measures on all finite cylinders, hence $P=Q$.

\section{Proof of Converse for Theorem \ref{thm: asymp opt rate and dec}}\label{App: pf of best rate}
Let $P_e=\Pr(\hatM\ne M)=\bar{\beta}(\gamma, (\bbP_k)_{k\in[m]})$. 
From the Fano's inequality, we have
\begin{equation}
    \rvH(M|\hatM,\zeta^T)\leq \rvH(M|\hatM)\leq 1 +P_e\log m.
\end{equation}
With Assumption \ref{Ass: strong secrecy}, the entropy of $M$ is upper bounded by
\begin{align}
    \log m = \rvH(M)=\rvH(M|\zeta^T)&=\rvI(M;\hatM|\zeta^T)+\rvH(M|\hatM,\zeta^T)\\
    &\leq \rvI(M;X^T|\zeta^T)+1 +P_e\log m\\
    &\leq H(X^T|\zeta^T)+1 +P_e\log m,
\end{align}
which leads to
\begin{align}
    \frac{\log m}{T}\leq \frac{H(X^T|\zeta^T)}{T}+\frac{1}{T}+P_e\frac{\log m}{T}.
\end{align}
If $P_e\to 0$ as $T\to\infty$, we have
\begin{align}
    \frac{\log m}{T} \leq \frac{H(X^T|\zeta^T)}{T}\leq \frac{\rvH(X^T)}{T}\leq \max_{P_{X^T}:\frac{1}{T}\rvD_\KL(P_{X^T}\|Q_{X^T})\leq d}\frac{\rvH(X^T)}{T}
\end{align}
and
\begin{equation}
    \limsup_{T\to\infty}\frac{\log m}{T} \leq \limsup_{T\to\infty}\max_{P_{X^T}:\frac{1}{T}\rvD_\KL(P_{X^T}\|Q_{X^T})\leq d}\frac{\rvH(X^T)}{T}=\max_{P:\overline\rvD_\KL(P\|Q)\leq d}\rvH_P(\calX),
\end{equation}
where the last inequality holds as the limit exists for a stationary ergodic process.

\section{Proof of Achievability for Theorem \ref{thm: asymp opt rate and dec}}\label{App: pf of thm: asymp opt rate and dec}

\paragraph{Embedding and decoding construction}
In the following, we identify the asymptotically optimal watermarking scheme that achieves vanishing $j$-th error probabilities and the maximum information rate, while satisfying the worst-case false-alarm error and distortion constraint. %Leveraging the asymptotic equipartition property (AEP), we  present the optimal design when distortion $d=0$, i.e., $P_{X^T}=Q_{X^T}$, as follows. Here, we use $\doteq$ to denote equality to the first order in the exponent.
Under any hypothesis $\rmH_j$ and any $\bbP_j$, we define the typical sets of stationary ergodic processes $x^T\in\calX^T$ and $\zeta^T\in\calZ^T$.
% \vspace{-1em}
\begin{definition}[Typical Sets]
    For arbitrarily small $\eta> 0$, 
    % define the set $\calA_{\eta,j}^{(T)}$ of jointly typical sequences $\{(x^T,\zeta^T)\}$ w.r.t.\ the distribution  $P_{X,\zeta|M=j}$ as
    % \begin{align}
    %     &\calA_{\eta,j}^{(T)}\coloneqq\bigg\{(x^T,\zeta^T)\in \calX^T\times \calZ^T: \nn\\
    %     &\bigg|\!\frac{\log P_X^T(x^T)}{T}-\! \rvH(X) \! \bigg|\! \leq \eta, \! \bigg|\! \frac{\log P_\zeta^T(\zeta^T)}{T}-\! \rvH(\zeta) \!\bigg|\! \leq \eta, \nn\\
    %     & \bigg|\frac{\log P_{X,\zeta|M=j}^T(x^T,\zeta^T)}{T}-\rvH(X,\zeta|M=j) \bigg|\leq \eta \bigg\}.
    % \end{align}
    define the  typical sets $\calA_{\eta,X}^{(T)}$ and $\calA_{\eta,\zeta}^{(T)}$ as
    \begin{align}
        \calA_{\eta,X}^{(T)}&\coloneqq \!\bigg\{x^T \!\!: \!\!\bigg|\frac{1}{T}\log \frac{1}{P_{X^T}(x^T)}-\rvH_P(\calX) \bigg| \!\! \leq \eta \!\bigg\},
        \calA_{\eta,\zeta}^{(T)}\coloneqq\bigg\{\zeta^T \!\!:  \!\!\bigg|\frac{1}{T}\log \frac{1}{P_{\zeta^T}(\zeta^T)}-\rvH_P(\calZ) \!\bigg|\leq \eta\bigg\}.
    \end{align}
\end{definition}
The typical sequences in $\calA_{\eta,X}^{(T)}$ and $\calA_{\eta,\zeta}^{(T)}$ are nearly uniformly distributed and can be mapped with almost deterministic precision. Leveraging the asymptotic equipartition property (AEP), we first present the optimal design when distortion $d=0$, i.e., $P_{X^T}=Q_{X^T}$, as follows. Here, we use $\doteq$ to denote equality to the first order in the exponent.

\paragraph{Step 1 (Embedding).}  Let  $\calZ\subset \bbZ$ and design a stationary ergodic process $P_\zeta^*$ on $\calZ$ such that $\rvH_P(\calZ)=\rvH_Q(\calX)$. Let $\eta=T^{-\frac{1}{4}}$. The  asymptotically optimal joint distribution $P_{X^T,\zeta^T|M}^*$ is constructed as follows: for any $i\in[m]$, %and arbitrarily small $\delta>0$:
\begin{itemize}[leftmargin=*,topsep=0pt]
    % \item Choose any $P_X^*$ such that $\sD(P_X^{*T},Q_X^T)\leq d$.
    % \item Let $\calZ\subset \bbZ$ and design $P_\zeta^*\in\calP(\calZ)$ such that $\rvH(\zeta)=\rvH(X)$. 
    \item for all $x^T\in \calA_{\eta,X}^{(T)}$, 
    $P_{X,\zeta|M}^{*T}(x^T,\zeta^T|i)=\begin{cases}Q_{X^T}(x^T) \doteq e^{-T\rvH_Q(\calX)}, & \!\!\!\! \text{if $\zeta^T \!\!\in \!\! \calA_{\eta,\zeta}^{(T)}$ and $g(x^T,\zeta^T)= i$};\\
    0, & \!\!\!\!\text{otherwise,} 
    \end{cases}$
    for some function $g:\calA_{\eta,X}^{(T)} \times \calA_{\eta,\zeta}^{(T)}\to \Big[|\calA_{\eta,X}^{(T)}|\Big]$ 
         such that it is bijective in either argument when the other is fixed.
    
    \item for all $x^T\notin \calA_{\eta,X}^{(T)}$, let $P_{X,\zeta|X,M}^{*T}(x^T,\zeta^T|i)$  take any non-negative value as long as $P_{X,\zeta|X,M}^{*T}$ is a valid probability distribution. %$\sum_{x^T,\zeta^T}P_{X,\zeta|X,M}^{*T}(x^T,\zeta^T|i)=1$.
\end{itemize}

\paragraph{Step 2 (Decoding).} The optimal decoder accepts the form:
    \setlength{\abovedisplayskip}{5pt}
    \setlength{\belowdisplayskip}{-1pt}
    \setlength{\abovedisplayshortskip}{5pt}
    \setlength{\belowdisplayshortskip}{-1pt}
\begin{align}
    % &\Gamma_\eta^*\coloneqq \!\!\Bigg\{ \!\!\gamma \Bigg| 
    \gamma^*(x^T,\zeta^T)\!=\!\!\begin{cases}
        g(x^T,\zeta^T), &\!\!\! \forall x^T\in \calA_{\eta,X}^{(T)}, \zeta^T \in \calA_{\eta,\zeta}^{(T)},  \text{ and } g(x^T,\zeta^T)\leq m;\\
        0, &\text{otherwise}.
        \end{cases}
\end{align}
\begin{figure}[t]
    \centering
    \includegraphics[width=0.6\linewidth]{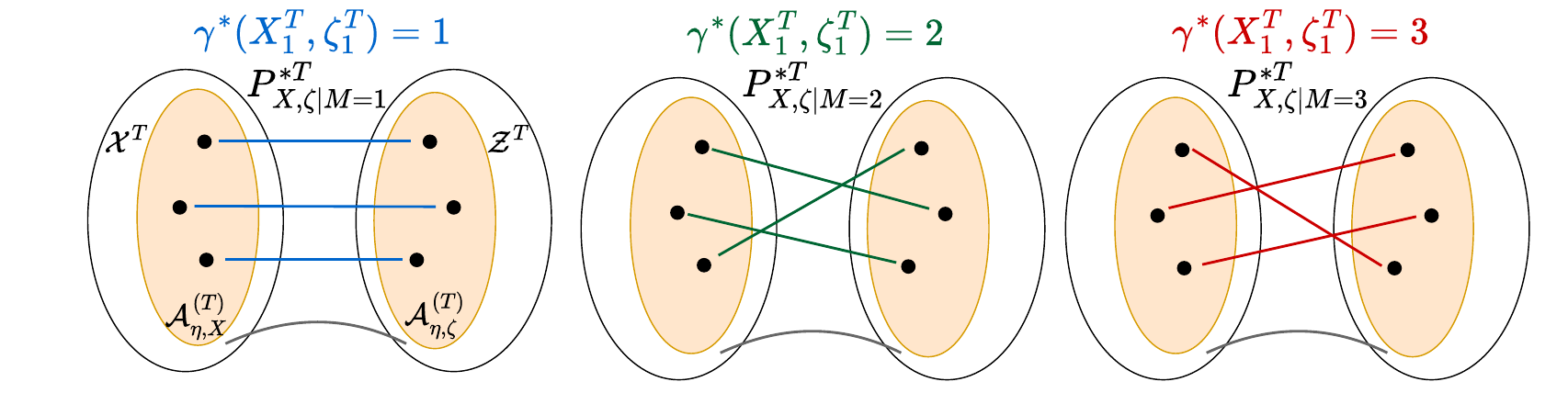}
    \vspace{-1em}
    \caption{Illustration of the asymptotically optimal watermarking scheme when $m = 3$.}
    \label{fig:asymp opt}
    \vspace{-1em}
\end{figure}
 Figure \ref{fig:asymp opt} illustrates the design using a toy example when $m=3$.

\paragraph{Existence of asymptotically optimal decoders}
First, the function $g$ proposed in Theorem \ref{thm: asymp opt rate and dec} always exists. For example, a Latin square matrix satisfies the structure.
If the message set size $m$ satisfies $\frac{1}{T}(\log m-\log \alpha)\leq \rvH_Q(\calX)$, then we have
\begin{equation}
    m \dotleq e^{T \rvH_Q(\calX)} \doteq |\calA_{\eta,X}^{(T)}|,
\end{equation}
and the output space of $g$ contains $[m]$. Thus, any optimal decoder $\gamma^*$ can decode messages drawn from $[m]$.

\paragraph{Asymptotic optimality} 
For any optimal decoder $\gamma^*$, one can always construct the corresponding joint distribution $P_{X,\zeta|M}^*$ in Theorem \ref{thm: asymp opt rate and dec}. In the following, we first show that the probability of the atypical set decays exponentially with $T$. We then prove that the $j$-th error probability vanishes to $0$ while the worst-case false-alarm error is upper bounded by $\alpha$ as $T\to\infty$.

Let $\eta=T^{-\frac{1}{4}}$ and define the set $\calA_{\eta,j}^{(T)}$ of jointly typical stationary ergodic processes $\{(x^T,\zeta^T)\}$ w.r.t.\ the distribution  $P_{X^T,\zeta^T|M=j}$ as
    \begin{align}
        \calA_{\eta,j}^{(T)}\coloneqq\bigg\{(x^T,\zeta^T)\in \calX^T\times \calZ^T: &\bigg|-\frac{1}{T}\log P_{X^T}(x^T)- \rvH_P(\calX)  \bigg| \leq \eta , \nn\\
        &\bigg|-\frac{1}{T}\log P_{\zeta^T}(\zeta^T)- \rvH_P(\calZ) \bigg| \leq \eta, \\
        & \bigg|-\frac{1}{T}\log P_{X^T,\zeta^T|M=j}(x^T,\zeta^T)-\rvH_P(\calX\times\calZ) \bigg|\leq \eta \bigg\}.
    \end{align}

First, we bound the probability of the atypical sets $(\calA_{\eta,X}^{(T)})^\rmc,(\calA_{\eta,\zeta}^{(T)})^\rmc,(\calA_{\eta,j}^{(T)})^\rmc$.
From the union bound, we have
\begin{align}
    \bbP_j((X^T,\zeta^T)\notin \calA_{\eta,j}^{(T)})&\leq \bbP_j\bigg(\bigg|-\frac{1}{T}\log P_{X^T}(x^T)- \rvH_P(\calX)  \bigg| \geq  \eta \bigg)\\
    &\quad +\bbP_j\bigg(\bigg|-\frac{1}{T}\log P_{\zeta^T}(\zeta^T)- \rvH_P(\calZ) \bigg| \geq  \eta \bigg)\\
    &\quad +\bbP_j\bigg(\bigg|-\frac{1}{T}\log P_{X^T,\zeta^T|M=j}(x^T,\zeta^T)-\rvH_P(\calX\times\calZ) \bigg| \geq  \eta \bigg). \label{Eq: atypical union bd}
\end{align}
Then, by the Chernoff bound, we have
\begin{align}
    &\bbP_j\bigg(\bigg|-\frac{1}{T}\log P_{X^T}(x^T)- \rvH_P(\calX) \bigg| \geq  \eta \bigg)\leq 2\bbP_j\bigg(\bigg|-\frac{1}{T}\log P_{X^T}(x^T)- \rvH_P(\calX)  \geq  \eta \bigg)\\
    &\leq 2\exp\bigg(-T\sup_{s\geq 0}(s\eta-\log \bbE[\exp(-s\log P_{X^T}(X^T))]) \bigg)\\
    &\overset{\text{(a)}}{\approx} 2\exp\bigg(-T\sup_{s\geq 0}(s\eta-\big(-s\bbE[\log P_{X^T}(X^T)]+s^2\bbE[(\log P_{X^T}(X^T))^2] \big) \bigg)\\
    &\eqb 2\exp(-\Omega(T\eta^2))=\exp(-\Omega(T^{\frac{1}{2}})),
\end{align}
where (a) follows from the Taylor expansion of $\exp(\cdot)$ and $\log(\cdot)$ and (b) follows since the maximum is achieved by $s=O(\eta)$.
The rest of the terms in the union bound \eqref{Eq: atypical union bd} can be similarly proved.

Thus, the probability of the jointly atypical set is upper bounded by
\begin{align}
    \bbP_j((X^T,\zeta^T)\notin \calA_{\eta,j}^{(T)})\leq 3\exp(-\Omega(T^{\frac{1}{2}}))=\exp(-\Omega(T^{\frac{1}{2}})).
\end{align}

Next, we prove that the proposed watermarking scheme in Theorem \ref{thm: asymp opt rate and dec} achieves the asymptotic optimality.
%Let $P_X^*=Q_X$, $\calZ\subset \bbZ$ and design $P^*_\zeta\in\calP(\calZ)$ such that $\rvH(P_\zeta^*)=\rvH(P_X^*)$. 

% Under the watermarking scheme in Theorem \ref{thm: asymp opt rate and dec}, we have $\supp(P_{X,\zeta|M=i}^{*T})=\supp(P_{X,\zeta|M=j}^{*T})$ for $i\ne j$, which ensure that $\min_{i:i\ne j}\sD_\KL(P^*_{X,\zeta|M=i}\|P^*_{X,\zeta|M=j})<\infty$ for any $P_X$. Let $$\eta\doteq e^{-TD^*}\coloneqq e^{-T\max_{P_X:\sD(P_X^T,Q_X^T)\leq d}\min_{i:i\ne j}\sD_\KL(P^*_{X,\zeta|M=i}\|P^*_{X,\zeta|M=j})}.$$

% For any $\gamma^*\in\Gamma^*$, any $j\in[m]$ and arbitrarily small $e^{T(\rvH(\zeta)-D^*)}\dotgeq\delta>0$, the $j$-th error probability is given by
Given any optimal $\gamma^*$ and the watermarking scheme  in Theorem \ref{thm: asymp opt rate and dec}, for any $j\in[m]$, the $j$-th error probability is given by
\begin{align}
    \beta_j(\gamma^*,P^*_{X^T,\zeta^T|M=j})&=\sum_{x^T,\zeta^T} P^*_{X^T,\zeta^T|M}(x^T,\zeta^T|j)\mathbbm{1}\{\gamma^*(x^T,\zeta^T)\ne j\}\\
    &\leq \sum_{(x^T,\zeta^T)\in \calA_{\eta,j}^{(T)}} P^*_{X^T,\zeta^T|M}(x^T,\zeta^T|j)\mathbbm{1}\{\gamma^*(x^T,\zeta^T)\ne j\} + \exp(-\Omega(T^{\frac{1}{2}}))\\
    &=\exp(-\Omega(T^{\frac{1}{2}})) \to 0 \text{ as } T\to \infty.
    % &\doteq \sum_{i\in[m]\backslash j}\sum_{(x^T,\zeta^T)\in \calA_{\eta,j}^{(T)}} \frac{\delta e^{-T\rvH(\zeta)}}{e^{T\rvH(\zeta)}-1}\mathbbm{1}\{\gamma^*(x^T,\zeta^T)=i\} + \eta\\
    % &\dotleq \delta e^{-T\rvH(\zeta)}+\eta\\
    % &\doteq \exp\big(-T\max_{P_X:\sD(P_X^T,Q_X^T)\leq d}\min_{i:i\ne j}\sD_\KL(P^*_{X,\zeta|M=i}\|P^*_{X,\zeta|M=j})\big).
\end{align}
% Thus, the $j$-th error probability achieves the exponential rate in Lemma \eqref{Lem:LB for j-th error}. 
For $j=0$, the worst-case false-alarm error probability is upper bounded as follows. For any $x^T\in \calA_{\eta,X}^{(T)}$,
\begin{align}
    &\sum_{\zeta^T}P^*_\zeta(\zeta^T)\mathbbm{1}\{\gamma^*(x^T,\zeta^T)\ne 0\}\\
    &\leq \sum_{\zeta^T\in\calA_{n,\zeta}^{(T)}}P^*_\zeta(\zeta^T)\mathbbm{1}\{\gamma^*(x^T,\zeta^T)\ne 0\} +\exp(-\Omega(T^{\frac{1}{2}}))\\
    &=\sum_{i\in[m]}\sum_{\zeta^T\in\calA_{n,\zeta}^{(T)}}P^*_\zeta(\zeta^T)\mathbbm{1}\{\gamma^*(x^T,\zeta^T)=i\} +\exp(-\Omega(T^{\frac{1}{2}}))\\
    &= \sum_{i\in [m]}\sum_{\zeta^T\in\calA_{n,\zeta}^{(T)}} \bigg(\frac{1}{m}\sum_{j\in[m]} \sum_{x^T}P^*_{X^T,\zeta^T|M}(x^T,\zeta^T|j) \bigg)\mathbbm{1}\{\gamma^*(x^T,\zeta^T)=i\} +\exp(-\Omega(T^{\frac{1}{2}}))\\
    &\doteq \sum_{i\in [m]}\sum_{\zeta^T\in\calA_{n,\zeta}^{(T)}}e^{-T\rvH(\calZ)}\mathbbm{1}\{\gamma^*(x^T,\zeta^T)=i\} +\exp(-\Omega(T^{\frac{1}{2}}))\\
    &= m e^{-T\rvH(\calZ)}+\exp(-\Omega(T^{\frac{1}{2}})) \\
    &\overset{\text{(a)}}{\leq} \alpha+\exp(-\Omega(T^{\frac{1}{2}}))\\
    &\xrightarrow{T\to\infty}\alpha,
\end{align}
where (a) follows from the condition $\log m \leq \log \alpha+T\rvH(\calZ)=\log \alpha+T\rvH_Q(\calX)$ in Theorem \ref{thm: asymp opt rate and dec}.

For any $x^T\in (\calA_{\eta,X}^{(T)})^\rmc$,
\begin{align}
    \sum_{\zeta^T}P^*_\zeta(\zeta^T)\mathbbm{1}\{\gamma^*(x^T,\zeta^T)\ne 0\}=0.
\end{align}
Since any distribution $Q_{X}^T$ can be written as a linear combinations of $\{\delta_{x^T}\}_{x^T\in\calX^T}$, we have
\begin{align}
    \sup_{Q_X}\beta_0(\gamma^*, Q_X \otimes P_\zeta^* )=\sup_{Q_X}\sum_{x^T,\zeta^T} Q_X^T(x^T)P^*_\zeta(\zeta^T)\mathbbm{1}\{\gamma^*(x^T,\zeta^T)\ne 0\}\to \alpha, \text{ as } T\to\infty.
\end{align}

\end{document}